\documentclass[a4paper, 12 pt] {article}
\usepackage{amsmath}
\usepackage{amsfonts}
\usepackage{amsthm}
\usepackage{amscd}

\newcommand\zvec{{\bf z}}
\newcommand\qvec{{\bf q}}
\newcommand\pvec{{\bf p}}

\newcommand\Rr{{\mathbb R}}
\newcommand\Tt{{\mathbb T}}

\newcommand\Zz{{\mathbb Z}}

\newcommand\wn{\hbox{\rm wn}\,}
\newcommand\codim{\hbox{\rm codim}\,}
\newcommand\sgn{\hbox{\rm sgn}\,}

\newcommand\corank{\hbox{\rm corank}\,}
\newcommand\Tr{\hbox{\rm Tr}\,}

\newcommand\cvec{{\bf c}}
\newcommand\Fvec{ F}

\newcommand\uvec{{\bf u}}
\newcommand\vvec{{\bf v}}
\newcommand\wvec{{\bf w}}

\newcommand\xvec{{\bf x}}

\newcommand{\half}{{\textstyle \frac{1}{2}}}
\newcommand{\fourth}{{\textstyle \frac{1}{4}}}

\newcommand\defined{\,{:=}\,}

\newcommand\Amat{{\mathrm A}}
\newcommand\Dmat{{\mathrm D}}
\newcommand\Fmat{{\mathrm F}}

\newcommand\Imat{{\mathrm I}}
\newcommand\Jmat{{\mathrm J}}
\newcommand\Kmat{{\mathrm K}}

\newcommand\Mmat{{\mathrm M}}
\newcommand\Mmatbar{{\overline \Mmat}}

\newcommand\Rmat{{\mathrm R}}
\newcommand\Rmatbar{{\overline \Rmat}}

\newcommand\Lmat{{\mathrm L}}
\newcommand\Lmatbar{{\overline \Lmat}}

\newcommand\Smat{{\mathrm S}}

\newcommand\Xvec{{\bf X}}

\newcommand\Acal{{\cal A}}
\newcommand\Abarcal{{\cal \overline A}}

\newcommand\Tcal{{\cal T}}
\newcommand\Tbarcal{{\cal \overline T}}

\newcommand\omegabar{{\overline \omega}}
\newcommand\taubar{{\overline \tau}}

\newcommand\cbar{{\overline c}}
\newcommand\cvecbar{{\bf \overline c}}

\newcommand\Gbar{{\overline G}}
\newcommand\ubar{{\overline u}}
\newcommand\uvecbar{{\bf \overline u}}

\newcommand\Ebar{{\overline E}}

\newcommand\Mbar{{\overline M}}
\newcommand\Lbar{{\overline L}}
\newcommand\Rbar{{\overline R}}
\newcommand\Tbar{{\overline T}}
\newcommand\lambdabar{{\overline \lambda}}
\newcommand\Lambdabar{{\overline \Lambda}}
\newcommand\gammabar{{\overline \gamma}}
\newcommand\nubar{{\overline \nu}}

\newcommand\xibar{{\overline \xi}}
\newcommand\etabar{{\overline \eta}}

\newcommand\contract{{\cdot}}

\newtheorem{thm}{Theorem}[section]

\newtheorem{prop}{Proposition}[section]
\newtheorem{cor}{Corollary}[section]

\numberwithin{equation}{section}

\numberwithin{equation}{section}
\begin{document}

\title{Singularities, Lax degeneracies and Maslov indices of the
  periodic Toda chain}

\author{JA Foxman$^\dag$ and JM Robbins\footnote{E-mail address: {\tt
      j.robbins@bristol.ac.uk}}$^{\dag,\ddag}$\\
$\dag$ School of Mathematics\\ 
University of Bristol,
  University Walk, Bristol BS8 1TW, UK\\
 $\ddag$ The Mathematical
  Sciences Research Institute\\ 1000 Centennial Drive, \#5070,
  Berkeley, CA 94720-5070, USA}

\thispagestyle{empty}

\maketitle

\begin{abstract}
  The $n$-particle periodic Toda chain is a well known example of an
  integrable but nonseparable Hamiltonian system in $\Rr^{2n}$.  We
  show that $\Sigma_k$, the $k$-fold singularities of the Toda chain,
  ie points where there exist $k$ independent linear relations amongst
  the gradients of the integrals of motion, coincide with points where
  there are $k$ (doubly) degenerate eigenvalues of representatives
  $\Lmat$ and $\Lmatbar$ of the two inequivalent classes of Lax
  matrices (corresponding to degenerate periodic or antiperiodic
  solutions of the associated second-order difference equation).  The
  singularities are shown to be nondegenerate, so that $\Sigma_k$ is a
  codimension-$2k$ symplectic submanifold. $\Sigma_k$ is shown to be
  of elliptic type, and the frequencies of transverse oscillations
  under Hamiltonians which fix $\Sigma_k$ are computed in terms of
  spectral data of the Lax matrices.  
  
  If $\mu(C)$ is the (even) Maslov index of a closed curve $C$ in the
  regular component of $\Rr^{2n}$, then $(-1)^{\mu(C)/2}$ is given by
  the product of the holonomies (equal to $\pm 1$) of the even- (or
  odd-) indexed eigenvector bundles of $\Lmat$ and $\Lmatbar$.
\end{abstract}

\section{Introduction}\label{sec:intro}

The Toda chain is a canonical example of a nonseparable but
integrable Hamiltonian system.  It consists of $n$ particles on the line with
exponential interactions between consecutively labeled particles.  In
the periodic Toda chain, the $n$th particle is coupled to the first.
The Hamiltonian is 
\begin{equation}
  \label{eq:H}
  H = \sum_{j=1}^n \half (p_j^2 + b_j^2),
\end{equation}
where 
\begin{equation}
  \label{eq:b_j}
  b_j = e^{(q_j - q_{j+1})/2},
\end{equation}
and $q_{n+1} \equiv q_1$.  

The Toda lattice was introduced in \cite{Toda}.  Its integrability was
established by H\'enon \cite{henon} and Flaschka \cite{flaschka} using
the method of Lax pairs.  There now exists an extensive literature on
the problem (see, eg, \cite{flaschka2}).  Properties of 
eigenvectors of Lax matrices over their associated spectral curve were studied in classical papers by Adler
and van Moerbeke \cite{adlvm1, adlvm2, adlvm3} and van Moerbeke and
Mumford \cite{vmmum}.  Audin \cite{audin} has used these results to
analyse the topology of the set of regular values of the integrals of
motion. A recent account is given by
Babelon, Bernard and Talon \cite{babelon}.

In the last 20 years there has been much interest in the topology of
integrable finite-dimensional Hamiltonian systems.  The generic local
structure and dynamics is given by the Liouville-Arnold theorem
\cite{arnoldcm}, according to which neighbourhoods of phase space are
foliated into invariant Lagrangian submanifolds diffeomorphic to
$\Rr^{n - k}\times\Tt^k$, where $\Tt^k$ is the $k$-torus, and the
dynamics is linearised by action-angle coordinates.  This local
behaviour breaks down at critical points of the energy-momentum map,
which comprise invariant sets of lower dimension.
A Morse theory for
integrable Hamiltonian systems, wherein the global topology is
described in terms of these critical sets, has been extensively
developed by Fomenko \cite{fomenko}, Eliasson \cite{eliasson}, Vey
\cite{vey} and Tien Zung \cite{tienzung}, among others.

Our interest here is in the phase space topology of the Toda chain.
The paper is organised as follows.  In Section~\ref{prop: offbanded},
we obtain Lax generators for the flows generated by a complete set of
$n$ integrals $\Fvec:\Rr^{2n}\rightarrow \Rr^n$.  These Lax generators
are constructed from representatives $\Lmat$ and $\Lmatbar$ of the two
inequivalent classes of symmetric Lax matrices for the periodic Toda
chain (these classes correspond to periodic and antiperiodic
solutions of the associated second-order difference equation).  Our
formulation is related to the classical treatment of van Moerbeke and
Mumford of general periodic finite difference operators \cite{vmmum},
but our approach is self-contained and quite elementary.
These
results are used to establish, in Section~\ref{sec:sing}, a one-to-one
correspondence between singularities of the Toda flow (critical points
of $\Fvec$) and eigenvalue degeneracies of the Lax matrices $\Lmat$
and $\Lmatbar$.  More precisely, the corank of $d\Fvec$ is equal to
the number of (doubly) degenerate eigenvalues of $\Lmat$ and
$\Lmatbar$.  Similar ideas relating singularities to points of degeneracy on
the spectral curve are discussed by Audin \cite{audin}.

In Section~\ref{sec: structure} we determine the local
structure of the corank-$k$ singularities $\Sigma_k$.  These are shown
to be codimension-$2k$ symplectic submanifolds of elliptic type
composed of ${n-1 \choose k}$ components disconnected from each other.
The eigenvalues of the linearised integrable flows which fix them are
computed in terms of the spectral data of $\Lmat$ and $\Lmatbar$.
There are singularities of corank between $1$ and $(n-1)$, and
$\Sigma_k$ is contained in the closure of $\Sigma_j$ for $k > j$. For
$n =3$, related results on the singularities and critical values for
the Toda chain and its algebraic generalisations have
been obtained by Polyakova \cite{polyakova} following the
programme of Fomenko \cite{fomenko}.  

The Maslov index is a topological invariant of Lagrangian tori of
integrable systems which appears in the semiclassical (EBK) quantisation
conditions  \cite{maslov, keller}.  In a companion paper \cite{FR1}, we show that the Maslov
index is determined by corank-one singularities.
In Section~\ref{sec: Maslov}, it is shown that the (even) Maslov index
of a closed curve in the set of regular points of $\Fvec$ is
determined, modulo 4, by the product of the even- (or odd-) indexed
holonomies of the eigenvector bundles of $\Lmat$ and $\Lmatbar$.  We
note that, because the Toda chain is nonseparable, application of the
semiclassical quantisation rules is not straightforward.

\section{Higher Lax flows}
The Lax formulation of the equations of motion for the Toda chain was 
obtained by Flaschka \cite{flaschka}.  Let $\Lmat$ be the
$n\times n$ symmetric matrix given by
 \begin{equation}
   \label{eq:L}
   \Lmat  = \left( 
       \begin{array}{rrrrrr}
           p_1&b_1&0&\cdots&0&b_n\\
           b_1&p_2&b_2&0&\cdots&0\\
   %        0&b_2&p_3&b_3&0&\cdots\\
           \vdots&&&&&\vdots\\
           0&\cdots&0& b_{n-2}&p_{n-1}&b_{n-1}\\
           b_n&0&\cdots&0&b_{n-1}&p_n
       \end{array}
 \right), 
\end{equation}
and $\Mmat_{(2)}$ the $n\times n$ antisymmetric matrix given by
\begin{equation}\label{eq:Mmat}
 \Mmat_{(2)} =  \frac{1}{2} \left( 
       \begin{array}{rrrrrr}
           0&b_1&0&\cdots&0&-b_n\\
           -b_1&0&b_2&0&\cdots&0\\
%           0&-b_2&0&b_3&0&\cdots\\
           \vdots&&&&&\vdots\\
           0&\cdots&0& -b_{n-2}&0&b_{n-1}\\
           b_n&0&\cdots&0&-b_{n-1}&0
       \end{array}
 \right).
 \end{equation}
(To simplify notion, in particular in Section~\ref{sec: structure}, 
we will sometimes write $\Mmat$ instead of
 $\Mmat_{(2)}$.)
 It is straightforward to verify that Hamilton's equations for the
 Hamiltonian (\ref{eq:H}) imply the Lax equation
\begin{equation}
  \label{eq:Lax equation}
  \dot \Lmat \defined \{\Lmat,H\} = [\Lmat,\Mmat_{(2)}].
\end{equation}
Conversely, (\ref{eq:Lax equation}) along with the (independent)
equation $\sum_{r=1}^n p_r = \sum_{r=1}^n \dot q_r$ imply Hamilton's
equations.  Thus,  Hamilton's equations and the Lax
equation are essentially equivalent.
It is convenient to express $\Lmat$ and $\Mmat_{(2)}$ in index form,
\begin{subequations}\label{eq:index form of L,M}
\begin{align}
  L_{rs} &= p_r\delta_{rs} +  b_r\delta_{r+1,s} + b_s\delta_{r,s+1},\label{eq:index form of L}\\
  M_{(2)rs} &=  b_r\delta_{r+1,s} -  b_s\delta_{r,s+1}\label{eq:index form of M}.
\end{align}
\end{subequations}
For convenience, here and elsewhere, matrices and vectors
are regarded as being periodic in their indices,
with period $n$.  Thus $L_{r+n,s} = L_{r,s+n} = L_{rs}$.
Similarly, we take $\delta_{rs}$ to be one if $r = s \mod n$ and to be zero
otherwise.

As is well known, the Lax equation implies that the eigenvalues of $\Lmat$, as well as
functions of them, are constants of the motion.  In particular, the
$n$ functions
\begin{equation}
  \label{eq:F_j}
  F_j = 
{{\textstyle \frac{1}{j}}} \,\Tr \Lmat^j, \quad 1 \le j \le n,
\end{equation}
are conserved.  We note that 
\begin{equation}
  \label{eq:F_1}
  F_1 = p_1 + \cdots + p_n
\end{equation}
is the centre-of-mass momentum, while $F_2$ is the Hamiltonian.  

In
this section we construct Lax equations for the Hamiltonian flows
generated by each of the $F_j$'s.  That is, we find antisymmetric
matrices $\Mmat_{(j)}$ such that
\begin{equation}
  \label{eq:higher Lax}
   \{\Lmat,F_j\} = [\Lmat,\Mmat_{(j)}]
\end{equation}
We may take $\Mmat_{(1)} = 0$ (since $F_1$ generates uniform translations
and $\Lmat$ is translation-invariant), while $\Mmat_{(2)}$ is given by
(\ref{eq:L}).  The formulation presented here is related to that of
van Moerbeke and Mumford \cite{vmmum}, who give nilpotent (triangular) Lax
generators for the higher flows of general periodic finite difference
operators.  Before giving expressions for $\Mmat_{(j)}$ for $j > 2$,
we recall that the higher Lax equations (\ref{eq:higher Lax}) already
imply that the functions $F_j$ are in involution.  Indeed,
\begin{equation}
  \label{eq:involution proof}
  \{F_{j+1},F_k\} = \Tr \left(\Lmat^j \{\Lmat,F_k\}\right) = \Tr
  \left(\Lmat^j [\Lmat,\Mmat_{(k)}]\right) = 0,
\end{equation}
where the last equality follows from the cyclicity of the trace.  We
recall, too, that integrability does not follow immediately from
(\ref{eq:involution proof}); one also needs to show that the $F_j$'s
are functionally independent (here, functional independence is implied
by Theorem~\ref{thm: symplectic submanifold} below).

We note that $H$ is invariant under the $n$ substitutions 
\begin{equation}
  \label{eq:substitutions}
  b_j
\mapsto -b_j.
\end{equation}
It follows that the Lax equation (\ref{eq:Lax equation}) is similarly
invariant under (\ref{eq:substitutions}).  It is easily seen that an
even number of such substitutions can be generated by conjugations
$\Lmat \mapsto \Smat\Lmat\Smat^{-1}$, $\Mmat_{(2)} \mapsto
\Smat\Mmat_{(2)}\Smat^{-1}$ for $\Smat$ a diagonal matrix of $\pm 1$'s. The
Lax equation (\ref{eq:Lax equation}) is trivially invariant under such
conjugations, as are the $F_j$'s.  There are, therefore, two
inequivalent classes of Lax pairs with respect to
(\ref{eq:substitutions}).  These are characterised by whether the
number of negative $b_j$'s is even or odd.  A spectral
characterisation of these even and odd classes is provided by the
difference equation
\begin{equation}
  \label{eq:diff eqn}
  b_r v_{r+1} + p_r v_r + b_{r-1}v_{r-1} = \lambda v_r.
\end{equation}
Eigenvalues $\lambda$ (with eigenvectors $\vvec$) of even $\Lmat$'s
correspond to periodic solutions $v_{r+n} = v_r$ of (\ref{eq:diff
  eqn}), whereas eigenvalues of odd $\Lmat$'s correspond to
antiperiodic solutions $v_{r+n} = -v_r$.
%WRONSKIAN???XXX??? IMPORTANT IN WHAT FOLLOWS
For definiteness and convenience, we take $\Lmat$, as given in
(\ref{eq:L}), to be the even representative, and $\Lmatbar$ to be
given by
replacing $b_n$ by $-b_n$, ie
\begin{equation}
  \Lbar_{rs} = p_r\delta_{rs} +  \sigma_r b_r\delta_{r+1,s} +
  \sigma_s b_s\delta_{r,s+1},\label{eq:index form of Lbar},
\end{equation}
where
\begin{equation}
  \label{eq:sigma_r}
  \sigma_r = 
  \begin{cases}
    -1,& r = 0 \mod n,\\
    \phantom{-}1, & \text{otherwise},
  \end{cases}
\end{equation}
to be the odd representative.

We say that an $n\times n$ symmetric matrix $\Amat$ is {\it off-banded
  of width $j$}, or {\it $j$-off-banded}, if it has precisely $(n-j)$
consecutive zero diagonals on or above main diagonal (thus, $j$ is the
number of diagonals above these zero diagonals, the first of which
does not vanish).  Equivalently, $\Amat$ is $j$-off-banded if
\begin{subequations}\label{eq:offbanded}
\begin{align}
  &A_{r, r+d} = 0,\ \ 1\le r \le n,  \ \ 0 \le d <  
\min(n-j-1,n-r),\label{eq:offbandeda}\\
  &\sum_{r=1}^j |A_{r, r+n - j}| \ne 0.\label{eq:offbandedb}
\end{align}
\end{subequations}
%We have the following:
\begin{prop}\label{prop: offbanded}
  For $1 \leq j \leq n$, $\Lmat^j - \Lmatbar^j$ is $j$-off-banded.
Moreover, its elements on the first nonzero diagonal are given by 
\begin{subequations}\label{eq: nonzero all j}
  \begin{align}
 %   \left(\Lmat^j\right)_{r, r+j} -   \left(\Lmatbar^j\right)_{r, r+j}
    \left(\Lmat^j - \Lmatbar^j\right)_{r, r+j}
&=   2 b_{r-1} b_{r-2} \cdots b_{r - j},
& j < n,
\label{eq: nonzero, j < n}\\
&= 4, & j = n, \label{eq: nonzero, j = n}
%  \left(\Lmat^j\right)_{r, r+j}   
%&=
%{\Lmat^n}_{rr} -  {\Lmatbar^n}_{rr} &= 4, & j = n. \label{eq: nonzero, j = n}
  \end{align}
\end{subequations}
where $1 \le r \le j$.
\end{prop}
%% (\ref{eq: nonzero, j < n}) implies that $\Tr (\Lmat^j - \Lmatbar^j) =
%% 0$ for $1\le j\le n$ while (\ref{eq: nonzero, j = n}) implies that
%% $\Tr (\Lmat^n - \Lmatbar^n) = 4n$, in accord with (\ref{eq:trace
%%   L^j-Lbar^j}).
\begin{proof}
Fix $1 \le r \le n$, and take $d$ such that $0 \le d \le \min(n-j-1,n-r)$.
From (\ref{eq:index form of Lbar}) and (\ref{eq:sigma_r}) it is clear that $(\Lmat^j - \Lmatbar^j)_{r,r+d}$ 
is given by twice the sum of terms of
  $(\Lmat^j)_{r,r+d}$ which are of odd degree in  $b_n$.  The terms of $(\Lmat^j)_{r,r+d}$
  are products $T(t_0,\ldots,t_j)$ of the form
  \begin{equation}
    \label{eq:T}
    T(t_0,\ldots,t_j) = L_{t_0t_1}L_{t_1 t_2} \cdots
  L_{t_{j-1}t_j},
  \end{equation}
where $t_0 = r$ and $t_j = r+d$.  $L_{1n}$ and
  $L_{n1}$, if they appear, contribute the only factors of $b_n$ to $
  T(t_0,\ldots,t_j)$.  
  
  Since $\Lmat$ is banded modulo $n$ (cf (\ref{eq:index form of L})),
  $T(t_0,\ldots,t_j)$ vanishes unless each pair of consecutive indices
  $t_k$ and $t_{k+1}$ in (\ref{eq:T}) differ by $0$ or $\pm 1$ modulo
  $n$.  Let us call the factor $L_{t_kt_{k+1}}$ a {\it right step} if
  $t_{k+1} = t_k + 1 \mod n$, and a {\it left step} if $t_{k+1} = t_k -1
  \mod n$ (diagonal factors, for which $t_{k+1} = t_k \mod n$, are
  neither left nor right steps).
  Let $u$ denote the number of right steps minus the number of left steps
  in $T(t_0,\ldots,t_j)$.  Then 
  \begin{equation}\label{eq: u =dmodn}
    u = d \mod n.
  \end{equation}
  Since $-j \le u \le j$ and, by assumption, $1 \le j \le n$ and $0
  \le d < n$, it follows that either $u = d$ or $u = d - n$.
  
  First, suppose that $u = d - n$.  We show that $T(t_0,\ldots,t_j)$
  is of odd degree in $b_n$.  This is certainly true for terms which
  contain no right steps.  Such terms are products of diagonal
  elements of $\Lmat$ (which do not contribute factors of $b_n$) and
  the left-step-only product
  \begin{equation}
    \label{eq:leftsteponly}
    L_{r,r-1}  L_{r-1,r-2} \cdots L_{r+(n-d)+1, r+(n-d)}.
  \end{equation}
%(by periodicity, the last factor in (\ref{eq:leftsteponly}) is just
%  $L_{r+d, r+d-1}).
%given modulo diagonal factors (which do not
%  contribute factors of $b_n$), $T(t_0,\ldots,t_j)$ by $L_{r,r-1}
%  L_{r-1,r-2} \cdots L_{21} L_{1n} L_{n,n-1} \cdots L_{r+d, r+d+1}$.
%  L_{r-1,r-2} \cdots L_{r+u, r+u-1}$.
  $b_n$ appears just once in (\ref{eq:leftsteponly}), in the factor
  $L_{10} := L_{1n}$.  A general term $T(t_0,\ldots,t_j)$ with $u =
  d-n$ is a product of diagonal elements, the left-step-only product
  (\ref{eq:leftsteponly}), and palindromic products of off-diagonal
  elements, ie, products of the form
  \begin{equation}
    \label{eq:subproduct}
     L_{k,k+1} L_{k+1,k+2} \cdots L_{k+p-1,k+p} \cdot L_{k+p,k+p-1} 
\cdots  L_{k+1,k+2} L_{k,k+1}
  \end{equation}
  in which every factor
   appears twice.
% (note that$ L_{k+p-1,k+p} = L_{k+p,k+p-1} $).
  
  Next, suppose that $d = u$.  A similar argument implies that
  $T(t_0,\ldots,t_j)$ is of even degree in $b_n$.  In this case, we
  note that the right-step-only product $L_{r,r+1} L_{r+1,r+2} \cdots
  L_{r+d-1,r+d}$ contains no factors of $L_{1n}$ or $L_{n1}$, and
  that, in general,
  $T(t_0,\ldots,t_j)$ is a product of diagonal elements, the
  right-step-only product, and palindromic products
  (\ref{eq:subproduct}).
  
  Thus, $T(t_0,\ldots,t_j)$ is of odd degree in $b_n$ if and only if $u
  = d - n$, or, equivalently, $d = n - |u|$.  Since $|u| \le j$, this condition can be satisfied
  only if $d \ge n - j$.  Thus, if $d < n - j$, 
  then $(\Lmat^j - \Lmatbar^j)_{r,r+d} = 0$,
  in accord with (\ref{eq:offbandeda}).
  
  To establish (\ref{eq:offbandedb}), we verify (\ref{eq: nonzero all
  j}), which implies that
  $\Lmat^j - \Lmatbar^j$ has nonzero elements on the $(n-j)$th
  diagonal above the main diagonal.  Let $d = n - j$.
 For $T(t_0,\ldots,t_j)$ to be of odd degree in $b_n$,  we require that
  $u = j$.  For $j
  < n$, there is only one such term, namely the left-step-only product
  $L_{r,r-1} L_{r-1,r-2}\cdots L_{r-j+1,r-j} = b_{r-1}
  b_{r-2}\cdots b_{r-j}$. (\ref{eq: nonzero, j < n}) follows.  For $j
  = n$, there are two nonzero terms with $u = n$.  The first is the
  left-step-only product $L_{r,r-1} L_{r-1,r-2}\cdots L_{r-n+1,r-n} =
  b_{r-1} b_{r-2} \cdots b_{r-n} = 1$.  The second is the
  right-step-only product $L_{r,r+1} L_{r+1,r+2}\cdots L_{r+n-1,r+n} =
  b_{r+1} b_{r+2} \cdots b_r = 1$.  (\ref{eq: nonzero, j = n})
  follows.
\end{proof}
From  Proposition \ref{prop: offbanded} it follows immediately that
  \begin{equation}\label{eq:TrL,TrLbar}
    \Tr \Lmat^j = 
    \begin{cases}
      \Tr \Lmatbar^j,& 1 \le j < n,\\
      \Tr \Lmatbar^n + 4n, & j = n.
    \end{cases}
  \end{equation}
Also, we note that if a linear combination of off-banded matrices vanishes and
the matrices all have different widths, then the coefficient of each
matrix vanishes (argue inductively, starting with the matrix of
greatest width).  Therefore, Proposition \ref{prop: offbanded} also implies the
following:
\begin{cor}\label{cor: lc of L^j - Lbar^j}
  If $\sum_{j=2}^n c_j (\Lmat^{j-1} - \Lmatbar^{j-1}) = 0$, then $c_2
  = \cdots = c_n = 0$.  
%Similarily, if $\sum_{j=2}^n c_j (\Mmat^{j-1}
%  - \Mmatbar^{j-1}) = 0$, then $c_2 = \cdots = c_n = 0$.
\end{cor}
 
We introduce the following notation: given
an $n\times n$ matrix $\Amat$, let $\Amat_+$ denote its strictly upper
triangular part, ie the matrix elements $A_{r,r+d}$ with $1\le r
\le n$ and $1 \le d \le n -r$. The generators of the Lax flows
for the $F_j$'s are given by the following:
\begin{prop}\label{prop: higher Laxes}
  Let $\Mmat_{(j)}$ and $\Mmatbar_{(j)}$ be the antisymmetric matrices
  given by
  \begin{subequations}\label{eq: defn of M_j and Mbar_j}
\begin{align}
    \Mmat_{(j)+}    &=\half (\Lmatbar^{j-1})_+,\label{eq: defn of M_j}\\
    \Mmatbar_{(j)+} &=\half (\Lmat^{j-1})_+\label{eq: defn of Mbar_j}
\end{align}
  \end{subequations}
for $1\le j \le n$.  Then
\begin{subequations}\label{eq: Lax eqns in prop}
\begin{align}
 \{\Lmat,F_j\}    &=   [\Lmat,\Mmat_{(j)}],   \label{eq: Lax for [L,F_j]}\\
 \{\Lmatbar,F_j\} &=   [\Lmatbar,\Mmatbar_{(j)}].\label{eq: Lax for [Lbar,F_j]}
\end{align}
\end{subequations}
\end{prop}
We note that $\Mmat_{(1)}$ and $ \Mmatbar_{(1)}$ both vanish
(consistent with the fact that $\Lmat$ is invariant under uniform
translations) while  for $j = 2$, (\ref{eq: defn of M_j}) agrees with
(\ref{eq:L}). 
\begin{proof}
  First, we note that (\ref{eq:TrL,TrLbar}) implies that under
  the substitution $b_n \mapsto -b_n$, $F_j$ changes by at most a constant.
  Therefore, the two sets of Lax equations equations (\ref{eq: Lax for
    [L,F_j]}) and (\ref{eq: Lax for [Lbar,F_j]}) are related by this
  substitution, and it suffices to verify just one of them.  For
  definiteness we consider (\ref{eq: Lax for [Lbar,F_j]}).
  
  Since both sides are symmetric matrices, it suffices to verify
  (\ref{eq: Lax for [Lbar,F_j]}) for elements on or above the main
  diagonal.  The elements of the left-hand side are given by
  \begin{equation}
    \label{eq:lhs of higher lax}
    \{\Lbar_{rs},F_j\} = \frac{1}{j} \Tr \{\Lbar_{rs}, \Lmat^j\} = \Tr
    \left(\{\Lbar_{rs}, \Lmat\} \Lmat^{j-1}\right).
  \end{equation}
From (\ref{eq:index form of L}), (\ref{eq:index form of Lbar}) and 
\begin{equation}
  \label{eq:pb b_r,p_r}
  \{b_r,p_s\} = \half b_r (\delta_{rs} - \delta_{r+1,s}),
\end{equation}
a straightforward calculation gives, for $1 \le r \le s \le n$, that
\begin{equation}
  \label{eq:lhs_2nd}
   \{\Lbar_{rs},F_j\} =
   \begin{cases}  b_{r-1}  (\Lmat^{j-1})_{r-1,r} - b_{r}  (\Lmat^{j-1})_{r,r+1},
& r =
  s,\\
 \half b_r \left( (\Lmat^{j-1})_{rr} -
  (\Lmat^{j-1})_{r+1,r+1}\right),& r+1 = s,\\
 \half b_n \left( (\Lmat^{j-1})_{11} - (\Lmat^{j-1})_{nn}\right),& r = 1, s
 = n,\\
0,& \text{otherwise}.
   \end{cases}
\end{equation}

Next we evaluate  the right-hand side of (\ref{eq: Lax for
  [Lbar,F_j]}), ie $[\Lmatbar, \Mmatbar_{(j)}]$.  From (\ref{eq:index form of Lbar}),
\begin{equation}
  \label{eq:rhs1}
   [\Lmatbar,\Mmatbar_{(j)}]_{rs} = p_r  \Mbar_{(j)rs} + \sigma_{r-1}
   b_{r-1} \Mbar_{(j)r-1,s} + \sigma_r b_r  \Mbar_{(j)r+1,s} +
   (r\leftrightarrow s).
\end{equation}
For $r = s$, (\ref{eq:rhs1}) yields 
$b_{r-1} {\Lmat^{j-1}}_{r-1,r} - b_r  {\Lmat^{j-1}}_{r+1,r}$,
%\begin{equation}
%  b_{r-1} {\Lmat^{j-1}}_{r-1,r} -
%b_r  {\Lmat^{j-1}}_{r+1,r},
%\end{equation}
in agreement with~(\ref{eq:lhs_2nd})
(note that $ \sigma_{r-1} \Mbar_{(j)r-1,r} = \half {\Lmat^{j-1}}_{r-1,r}$,
and similarly, $\sigma_r  \Mbar_{(j)r+1,r} = -\half{\Lmat^{j-1}}_{r+1,r}$).

To evaluate the off-diagonal elements $1 \le r < s \le n$ in
 (\ref{eq:lhs_2nd}), we will make use of the
  trivial identity $ [\Lmat, \half \Lmat^{j-1}] = 0$, or
  \begin{equation}
    \label{eq:ident components}
     [\Lmat,\half \Lmat^{j-1}]_{rs} = \half p_r  (\Lmat^{j-1})_{rs} +
     \half b_{r-1} (\Lmat^{j-1})_{r-1,s} +  \half b_{r}
     (\Lmat^{j-1})_{r+1,s} - (r\leftrightarrow s) = 0.
  \end{equation}
  Referring to
  the terms in (\ref{eq:rhs1}), we  use (\ref{eq: defn of Mbar_j}) to express $\Mmat_{j}$
in terms of $\Lmat^{j-1}$, as follows:
  \begin{align}\label{eq:rhs2}
    p_r \Mbar_{(j)rs} + (r \leftrightarrow s) &= p_r (\half
    \Lmat^{j-1})_{rs} - (r \leftrightarrow s),\nonumber\\
   \sigma_{r-1} b_{r-1}\Mbar_{(j)r-1,s} + (r \leftrightarrow s) &= b_{r-1} (\half
    \Lmat^{j-1})_{r-1,s} - (r \leftrightarrow s)\nonumber\\
&\ \ -\delta_{r1} \delta_{sn} b_n (\half \Lmat^{j-1})_{nn} +
    \delta_{r+1,s}b_r (\half \Lmat^{j-1})_{rr},\nonumber\\
 \sigma_{r} b_r\Mbar_{(j)r+1,s} + (r \leftrightarrow s) &= b_r (\half
    \Lmat^{j-1})_{r+1,s} - (r \leftrightarrow s)\nonumber\\
&\ \ +\delta_{r1} \delta_{sn} b_n (\half \Lmat^{j-1})_{11} -
    \delta_{r+1,s}b_r (\half \Lmat^{j-1})_{r+1,r+1}.
  \end{align}
Substituting the preceding into (\ref{eq:rhs1}) and using the identity
(\ref{eq:ident components}), we get
that
\begin{multline}
  \label{eq:rhs3}
  [\Lmatbar,\Mmatbar_{(j)}]_{rs} = [\Lmat,\half\Lmat^{j-1}]_{rs}
  +
  \half \delta_{r+1,s} b_r ((\Lmat^{j-1})_{rr} -
  (\Lmat^{j-1})_{r+1,r+1}) + \\ \half b_n \delta_{r1}\delta_{sn}
((\Lmat^{j-1})_{11} -
  (\Lmat^{j-1})_{nn}).
\end{multline}
As the first term vanishes, this agrees with (\ref{eq:lhs_2nd}).
\end{proof}

\section{Singularities and eigenvalue degeneracies}\label{sec:sing}

The singularities of an integrable system
$\Fvec:\Rr^{2n}\rightarrow \Rr^n$ are the critical points of $\Fvec$.
Here we show that singularities of the Toda chain coincide with
eigenvalue degeneracies of the Lax matrices $\Lmat$ and $\Lmatbar$.
Let $\Sigma$ denote the set of
singularities of the Toda chain, and 
let $\Sigma_k \subset \Sigma$ denote the subset in
which there are precisely $k$ linear relations amongst the $dF_j$'s, ie
\begin{equation}
  \label{eq:Sigma_k}
  \Sigma_k = \{(\qvec,\pvec) |\, \corank d\Fvec = k\}.
\end{equation}
We observe that eigenvalues of $\Lmat$ and $\Lmatbar$ are at most
two-fold degenerate, since the associated eigenvectors are solutions
of the second-order linear difference equation (\ref{eq:diff eqn}),
which for a given value of $\lambda$ has at most two linearly
independent solutions.  Let $\nu$ and $\nubar$ denote the number of
doubly degenerate eigenvalues of $\Lmat$ and $\Lmatbar$
respectively.

%% HERE IS

\begin{thm}\label{thm: sing and degen}
\begin{equation}
  \corank d\Fvec = \nu + \nubar.
\end{equation}
\end{thm}
\begin{proof}
  Let $\zvec_* \in \Rr^{2n}$. For convenience, let $\Fvec_*$ denote
$\Fvec(\zvec_*)$, $\Lmat_*$ denote $\Lmat(\zvec_*)$, etc.
%fin phase space, and for
%  notational convenience, we regard all functions on phase space,
%  when denoted without arguments, as being evaluated at $\zvec$ (eg,
%  $\Lmat$ will denote $\Lmat(\zvec_0)$ and
%  $dF_j$ will denote $dF_j(zvec_0)$).
%For example, $\nu$ denotes the number of
%  degenerate eigenvalues of $\Lmat$ at $\zvec$.  
Let $V_* \subset \Rr^n$ denote the space of linear relations amongst
the $dF_{j*}$'s, ie
  \begin{equation}
    \label{eq:V_set}
    V_* = \left\{\cvec \in \Rr^n  \Big | \,  \sum_{j=1}^n c_j dF_{j*} = 0 \right\}.
  \end{equation}
  so that $\corank d\Fvec_* = \dim V_*$. We show that
\begin{equation}
  \label{eq:tobeproved}
  \dim V_*
= \nu_* + \nubar_*.
\end{equation}

We first show that $\dim V_*\le \nu_* + \nubar_*$. 
%% \begin{equation}
%%   \label{eq:dimV<=nu+nubar}
%%    \dim V\le \nu + \nubar.
%% \end{equation}
Let $\Acal_*$ denote the real antisymmetric commutant of $\Lmat_*$;
that is, $\Acal_*$ consists of all $n$-dimensional real antisymmetric
matrices which commute with $\Lmat_*$.  For $\cvec \in V_*$,
Proposition~\ref{prop: higher Laxes} and Eq.~(\ref{eq:V_set}) imply that
\begin{equation}
  \label{eq:c_in_commutant}
  \left[\Lmat_*, \sum_{j=1}^n c_j \Mmat_{(j)*}\right] =\left \{\Lmat, \sum_{j=1}^n c_j
  F_{j} \right\}_* = 0,
\end{equation}
so that $\sum_{j=1}^n c_j \Mmat_{(j)*} \in \Acal_*$.  Similarly, letting $\Abarcal_*$
denote the real antisymmetric commutant of $\Lmatbar$, we have that
$\sum_{j=1}^n c_j
\Mmatbar_{(j)*} \in \Abarcal_*$.  Regarding $\Acal_*$ and $\Abarcal_*$ as
real vector spaces, we define a linear map from $V_*$ to 
$\Acal_*\oplus\Abarcal_*$ according to
\begin{equation}
  \label{eq:V_to_A+Abar}
  \cvec\mapsto \left(\sum_{j=1}^n c_j \Mmat_{(j)*}\right) \oplus \left(\sum_{j=1}^n c_j  \Mmatbar_{(j)*}\right).
\end{equation}
This map is 1-1, for
if $\sum_{j=1}^n c_j \Mmat_{(j)*} = \sum_{j=1}^n c_j \Mmatbar_{(j)*}
= 0$, then from (\ref{eq: defn of M_j and Mbar_j}) and 
Corollary~\ref{cor: lc of L^j - Lbar^j} it follows that 
\begin{equation}
  \label{eq:c_2=...}
  c_2 = \cdots = c_n = 0.
\end{equation}
But (\ref{eq:c_2=...}) and (\ref{eq:V_set})  imply
that $c_1 dF_1 = 0$.  Since $dF_1 \ne 0$ (cf (\ref{eq:F_1})),
we must have $c_1 = 0$, and
therefore $\cvec = 0$.  Thus, (\ref{eq:V_to_A+Abar}) is 1-1, and
\begin{equation}
  \label{eq:dimV<dimA+dimAbar}
  \dim V_* \le \dim \Acal_* + \dim \Abarcal_*.
\end{equation}

To compute $\dim \Acal_*$, we note that $\Acal_*$ is the
direct sum of the spaces of antisymmetric linear maps on the eigenspaces
of $\Lmat_*$ (endowed with the standard inner product from $\Rr^n$).  
In general, the space of antisymmetric linear maps on a
$k$-dimensional inner product space is of dimension $k(k-1)/2$.  Since
$\Lmat_*$ has $\nu_*$ two-dimensional eigenspaces and $(n-2\nu_*)$
one-dimensional eigenspaces, it follows that $\dim \Acal_* = \nu_*$.
Similarly, $\dim \Abarcal_* = \nubar_*$.  Substitution into
(\ref{eq:dimV<dimA+dimAbar}) yields 
\begin{equation}
  \label{eq:dim V<}
  \dim V_* \le \nu_* + \nubar_*.
\end{equation}

Next, we show that $\nu_* + \nubar_* \le \dim V_*$.  Let $\Tcal_*$ denote the
set of polynomials which annihilate $\Lmat_*$.
Elements of $\Tcal_*$ are divisible by the
minimum polynomial of $\Lmat_*$, which we denote by $P_*(x)$.
$P_*(x)$ is  of degree $n - \nu_*$.
Regarding $\Tcal_*$ as a vector space, we let $\Tcal_*^n$ denote the
subspace of polynomials of degree at most $n$, ie
\begin{equation}
  \label{eq:Tcal^n}
  \Tcal_*^n = \{R(x) | R(\Lmat_*) = 0, \deg R \le n  \}.
\end{equation}
Clearly
$\dim \Tcal_{*}^n = \nu_*$ (elements of $\Tcal_{*}^n$
are products of $P_*(x)$ with arbitrary
polynomials of degree at most $\nu_*$).  
Suppose $\sum_{j=1}^n c_j
x^{j-1} \in \Tcal_{n*}$.  Then 
\begin{equation}
  \label{eq:implication}
  \sum_{j=1}^n c_j {\Lmat_*}^{j-1} = 0 \implies
\Tr (\sum_{j=1}^n c_j {\Lmat_*}^{j-1}d\Lmat_*) = 0 \implies \sum_{j=1}^n c_j
dF_{j*} = 0.
\end{equation}
Therefore, 
\begin{equation}
  \label{eq:linear2}
  \sum_{j=1}^n c_j x^{j-1} \mapsto \cvec
\end{equation}
defines a linear map from $\Tcal_{*}^n$ to
$V_*$.  Clearly the map (\ref{eq:linear2}) is  1-1.
Similarly, let $\Tbarcal_*^n$ denote the $\nubar_*$-dimensional space of polynomials of
 degree at most $n$ which annihilate $\Lmatbar_*$, ie
 \begin{equation}
   \label{eq:Tbarcal_*^n}
   \Tbarcal_*^n = \{\Rbar(x) | \Rbar(\Lmat_*) = 0, \deg \Rbar \le n  \}.
 \end{equation}
 Arguing as above, we see that (\ref{eq:linear2}) also defines a 1-1
 linear map from $\Tbarcal_{*}^n$ to $V_*$.

Regarded as polynomial subspaces, $\Tcal_*^n$ and $\Tbarcal_*^n$ are
transverse.  For if $ \sum_{j=1}^n c_j x^{j-1} $ belongs to both, 
then 
$\sum_{j=1}^n c_j \Lmat^{j-1} = \sum_{j=1}^n c_j \Lmatbar^{j-1} = 0$.
Corollary~\ref{cor: lc of L^j - Lbar^j} implies that $c_2 = \cdots
= c_n = 0$, which in turn implies that $c_1 = 0$. 
Therefore, (\ref{eq:linear2}) defines a 1-1 map from $\Tcal_*^n\oplus
\Tbarcal_*^n$ to $V_*$, and
\begin{equation}
  \label{eq:greater_than}
  \nu_* + \nubar_*= \dim (\Tcal_*^n\oplus
\Tbarcal_*^n) \le \dim V_*,
\end{equation}
as required.
\end{proof}
Since we actually have an equality in (\ref{eq:greater_than}),
we deduce the following:
\begin{cor}\label{cor: t_* and V_*}
Given  
$\Tcal_*^n$, $\Tbarcal_*^n$ and $V_*$ as above,
\begin{equation}
  \label{eq:map again}
   \sum_{j=1}^n c_j x^{j-1} \mapsto \cvec
\end{equation}
is an isomorphism from $\Tcal_*^n\oplus \Tbarcal_*^n$ to $V_*$.
\end{cor}

\section{Structure of singular sets}\label{sec: structure}

The Toda Hamiltonian has no corank-$n$
singularities, since $dF_1 \ne 0$.
The singularities of corank $(n-1)$ are relative
equilibria, as is shown in the following:
\begin{prop}\label{prop: Omega_n}
  Let $\Omega_{n-1} = \{(\qvec,\pvec) |\, q_1 = \cdots = q_n, p_1 = \cdots
  p_n\}$ denote the set of points $(\qvec,\pvec)$ for which the
  components of $\qvec$ are all the same and the components of $\pvec$
  are all the same.  Then
\begin{equation}
  \label{eq:Sigma_2n-2}
  \Sigma_{n-1} = \Omega_{n-1}.
\end{equation}
\end{prop}
\begin{proof}
First we show that $\Sigma_{n-1} \subset \Omega_{n-1}$.  
Let $(\qvec,\pvec) \in \Sigma_{n-1}$.
Since $dF_1 \ne 0$, $dH(\qvec,\pvec)$ must be proportional to $dF_1$.  But 
\begin{equation}
  \label{eq:dH}
  dH = \sum_{j=1}^n p_j dp_j +  \sum_{j=1}^n \left(b_j^2 - b_{j-1}^2\right) dq_j.
\end{equation}
For $dH$ to be proportional to $dF_1$, we must have that the $p_j$'s
are all the same and the $b_j$'s (which are positive) are all the
same.  The latter implies that $q_{j+1} - q_j$ is a constant
independent of $j$, and periodicity, ie $q_{n+1} = q_1$, then implies
that the $q_j$'s are all the same.  

Next we show that $\Omega_{n-1}\subset \Sigma_{n-1}$.
For $(\qvec,\pvec) \in \Omega_{n-1}$, the difference equation (\ref{eq:diff
  eqn}) simplifies to
\begin{equation}
  \label{eq:degen difference}
  v_{r-1} + v_{r+1} = (\lambda - p) v_r,
\end{equation}
where $p$ is the common value of the components of $\pvec$.
Periodic and antiperiodic solutions of (\ref{eq:degen difference}) are given by 
\begin{align}
  \label{eq:periodic solutions}
 %%  u^{\pm}_{(r)j} &= \exp\left(\pm \pi i j \frac{r}{n}\right), 
%%    & \lambda_r = p + 2\cos\left(\pi  \frac{r}{n}\right),\ 
%% &0 \le  r \le n,\ &r\ \text{even},&\\
  u^{\pm}_{(r)j} &= \exp\left(\pm \pi i j r/n \right), 
   & \lambda_r = p + 2\cos\left(\pi  r/n\right),\ 
&0 \le  r \le n,\ &r\ \text{even},&\\
 \label{eq:antiperiodic solutions}
%%   \ubar^{\pm}_{(s)j} &= \exp\left(\pm \pi i j \frac{r}{n}\right), 
%%   & \lambdabar_s = p + 2\cos\left(\pi  \frac{s}{n}\right),\ 
%% &0 <  s \le n,\ &s\ \text{odd}.&
  \ubar^{\pm}_{(s)j} &= \exp\left(\pm \pi i j s/n\right), 
  & \lambdabar_s = p + 2\cos\left(\pi s/n\right),\ 
&0 <  s \le n,\ &s\ \text{odd}.&
\end{align}
The $\lambda_r$'s are doubly degenerate except for
$r$ equal to $0$ and (if $n$ is even) $n/2$, 
while the $\lambdabar_s$'s are all doubly degenerate.
Thus,
%%  for $n$ even, 
%% \begin{equation}
%%   \label{eq:n even}
%%   \nu =  \half n - 1,\quad \nubar  = \half n,
%% \end{equation}
%% while
%% for $n$ odd, 
%% \begin{equation}
%%   \label{eq:n odd}
%%   \nu = \half(n-1), \quad \nubar = \half(n-1).
%% \end{equation}
\begin{equation}
  \label{eq:n evenodd}
  \nu =  [\half (n - 1)], \quad \nubar  = [\half n],
\end{equation}
where $[x]$ denotes the integer part of $x$.
In general, $\nu + \nubar = n-1$, so Theorem~~\ref{thm: sing and
  degen} implies that $(\qvec,\pvec) \in \Sigma_{n-1}$.
\end{proof}

The explicit expressions (\ref{eq:periodic solutions}) and
 (\ref{eq:antiperiodic solutions}) which hold in  $\Sigma_{n-1}$
 allows us to deduce the 
 following general result:
\begin{prop}\label{prop: allowed degens}
  Let $\lambda_1 \ge \cdots \ge \lambda_n$ and $\lambdabar_1 \ge
  \cdots \ge \lambdabar_n$ denote the eigenvalues of $\Lmat$ and
  $\Lmatbar$ in descending order.  Then $\lambda_r > \lambdabar_r$ for
  $r$ odd, $\lambdabar_r > \lambda_r$ for $r$ even, and
 the allowed degeneracies
  are $\lambda_{r} = \lambda_{r+1}$ for $r$ even and
$\lambdabar_{r} = \lambdabar_{r+1}$ for $r$ odd.
\end{prop}
\begin{proof}
  From (\ref{eq:periodic solutions}) and (\ref{eq:antiperiodic
  solutions}), these degeneracies are simultaneously realised at the points of
  $\Sigma_{n-1}$.  Indeed, for points in $\Sigma_{n-1}$, we have that
%% \begin{align}\label{eq:Omega relations}
%%   \lambda_{2j-1} < \lambdabar_{2j-1} &=  \lambdabar_{2j} <
%%   \lambda_{2j},\\
%%  \lambdabar_{2j} < \lambda_{2j} &= \lambda_{2j+1} <
%%   \lambdabar_{2j+1}
%% \end{align}
\begin{equation}
  \label{eq:Omega relations}
 \lambda_1 > \lambdabar_1 = \lambdabar_2 > \lambda_2 = \lambda_3 >
 \cdots \lambdabar_{2j-1} = \lambdabar_{2j} > 
\lambda_{2j} = \lambda_{2j+1 } > \cdots
\end{equation}
(the form of the end of the sequence  depends on whether $n$ is even or
odd).  
%Thus, in this case, if the eigenvalues of $\Lmat$ and
%$\Lmatbar$ are arranged in descending order, the sequence consists
%alternating pairs from $\Lmat$ and $\Lmatbar$ (except for singletons
%from $\Lmat$ at the beginning and, if $n$ is even, the
%end).

For any $\zvec\in \Rr^{2n}$, (\ref{eq:TrL,TrLbar}) along with
Newton's identities  imply that the
characteristic polynomials of $\Lmat$ and $\Lmatbar$ differ by a constant:
\begin{equation}
  \label{eq:char poly diffs}
  \det(x -\Lmat(\zvec)) = \det(x -
\Lmatbar(\zvec)) + 4.
\end{equation}
Therefore, in general, eigenvalues of $\Lmat$ and $\Lmatbar$ cannot coincide.
Since the eigenvalues  depend continuously on $\zvec$, it
follows that the inequalities in (\ref{eq:Omega relations}) hold not
only in $\Sigma_{n-1}$ but everywhere else.
Therefore, in general, we have that
\begin{equation}
  \label{eq:not Omega relations}
 \lambda_1 > \lambdabar_1 \ge \lambdabar_2 > \lambda_2 \ge \lambda_3 >
 \cdots \lambdabar_{2j-1} \ge \lambdabar_{2j} > 
\lambda_{2j} \ge \lambdabar_{2j+1 } > \cdots
\end{equation}
It follows that $\lambda_{2j} = \lambda_{2j+1}$ and $\lambdabar_{2j-1} =
\lambdabar_{2j}$ are the only possible degeneracies.
\end{proof}

To determine the local structure of the singular set $\Sigma$, it is
convenient%, in a neighbourhood of one of its points, 
to bring the Lax
matrices $\Lmat$ and $\Lmatbar$ to a canonical form.  As above, let $\zvec_* \in
\Sigma_k$. 
%(below we show that $\Sigma_k$ is nonempty for $k =
%1,\ldots,n-1$).  
Let $\Lmat_*$ denote $\Lmat(\zvec_*)$, and let other functions
evaluated at $\zvec_*$ be similarly denoted.  Let $\nu_*$ and
$\nubar_*$ denote the number of (doubly) degenerate eigenvalues of
$\Lmat_*$ and $\Lmatbar_*$ respectively.  From Theorem~\ref{thm: sing
  and degen}, $\nu_* + \nubar_* = k$.  Let $\lambda_{r*}$, $1 \le r
\le n$ denote the eigenvalues of $\Lmat_*$ with degenerate eigenvalues
repeated, ordered so that $\lambda_{1*} = \lambda_{2*}, \ldots,
\lambda_{2\nu-1*} = \lambda_{2\nu*}$.  Let $\uvec_r$ denote an
orthonormal set of eigenvectors of $\Lmat_*$.  Define
$\lambdabar_{s*}$ and $\uvecbar_{s}$ similarily with respect to
$\Lmatbar_*$.  Then for $\zvec$ in some neighbourhood of $\zvec_*$,
there exists an orthogonal matrix $\Rmat(\zvec)$ depending smoothly on
$\zvec$, with $\Rmat_* = \Imat$, such that
$\Rmat^T(\zvec)\Lmat(\zvec)\Rmat(\zvec)$ is block diagonal with
respect to the $\uvec_{r}$-basis, with two-dimensional blocks for $1
\le r \le 2\nu_*$ and diagonal for $2 \nu_* < r \le n$.  %(The fact
%that the only degeneracies are those described by \ref{prop: allowed
%  degens} implies that $\Rmat$ can be defined everywhere.) 
That is,
letting $\Lambda(\zvec)$ be the symmetric matrix with elements
\begin{equation}
  \label{eq:Lambda elements}
  \Lambda_{rt}(\zvec) =
\uvec_{r}\cdot\Rmat^T(\zvec)\Lmat(\zvec)\Rmat(\zvec)\cdot\uvec_{t},
\end{equation}
we have that
\begin{equation}
  \label{eq:Lambda(zpvec)}
  \Lambda(\zvec) = 
  \begin{pmatrix}
%   1   2   3   4   5   6   7   8 
    * & * &   &   &   &   &   &   \\
    * & * &   &   &   &   &   &   \\
      &   &\ddots&   &   &   &   &   \\
      &   &   & * & * &   &   &   \\
      &   &   & * & * &   &   &   \\
      &   &   &   &   & * &   &   \\
      &   &   &   &   &   &\ddots& \\
      &   &   &   &   &   &   & * 
  \end{pmatrix},
\end{equation}
where omitted entries are zeros.
At $\zvec_*$,
\begin{equation}
  \label{eq:Lambda*}
\Lambda_* = 
  \begin{pmatrix}
%   1   2   3   4   5   6   7   8 
    \lambda_{1*} & 0 &   &   &   &   &   &   \\
   0 & \lambda_{1*} &   &   &   &   &   &   \\
      &   &\ddots&   &   &   &   &   \\
      &   &   & \lambda_{r*} & 0 &   &   &   \\
      &   &   & 0 & \lambda_{r*} &   &   &   \\
      &   &   &   &   & \lambda_{r+1*} &   &   \\
      &   &   &   &   &   &\ddots& \\
      &   &   &   &   &   &   & \lambda_{n-2\nu*} 
  \end{pmatrix}.
\end{equation}
Similarly, in some neighbourhood of $\zvec_*$,
there exists an orthogonal matrix $\Rmatbar(\zvec)$ such that
$\Rmatbar^T(\zvec)\Lmatbar(\zvec)\Rmatbar(\zvec)$ is block diagonal in
the $\uvecbar_{s*}$ basis, with two-dimensional blocks for $1 \le s
\le 2\nubar_*$ and diagonal for $2 \nu_* < s \le n$.  Define
$\Lambdabar(\zvec)$ in analogy with (\ref{eq:Lambda elements}), ie
\begin{equation}
  \label{eq:Lambdabar elements}
  \Lambdabar_{su}(\zvec) =
\uvecbar_{s}\cdot\Rmat^T(\zvec)\Lmat(\zvec)\Rmat(\zvec)\cdot\uvecbar_{u}.
\end{equation}
%HERE 

Let 
\begin{align}\label{eq: xi,eta}
   \xi_r & = \half(\Lambda_{2r, 2r} -  \Lambda_{2r-1, 2r-1}), & \eta_r & =
   \Lambda_{2r-1, 2r}, & 1 \le r \le \nu,\nonumber\\
   \xibar_s & = \half(\Lambdabar_{2s, 2s} -  \Lambdabar_{2s-1, 2s-1}), &\etabar_s & =
   \Lambdabar_{2s-1, 2s}, & 1 \le s \le \nubar.
\end{align}
Then, from (\ref{eq:Lambda elements}) and (\ref{eq:Lambdabar
  elements}), $\Sigma_k$ is given locally 
by the vanishing of $\xi_r$,
  $\eta_r$ and $\xibar_s$, $\etabar_s$, eg
\begin{equation}
  \label{eq:local_Sigma_k}
% \Sigma_k\cap U  = \{\zvec \in U\, \xi_r =
%  \eta_r = \xibar_s = \etabar_s = 0, 1 \le r \le \nu_*, 1 \le s \le
%  \nubar_*\}
% \Sigma_k\cap U  = \{\zvec \in U\, 
\xi_r =  \eta_r = 0,\ \ 1 \le r \le \nu_*, \quad
\xibar_s = \etabar_s = 0,\ \ 1 \le s \le \nubar_*
\end{equation}
in some neighbourhood of $\zvec_*$.

The next result, Proposition~\ref{prop: Poisson bracket of spectral
  components of L}, is a general expression for Poisson brackets of
spectral components of the Lax matrices evaluated at $\zvec_*$.  It is
used in Proposition~\ref{prop: xi,eta canonical} to show that the
Poisson brackets amongst $\xi_r$, $\eta_s$, $\xibar_s$ and $\etabar_s$
evaluated at $\zvec_*$ are, up to normalisation, of canonical form.
For the statement of Proposition~\ref{prop: Poisson bracket of
  spectral components of L},
% of \ref{prop: Poisson bracket of spectral components of L}, 
it is convenient to introduce Lax matrices in which
the $b_j$'s are allowed to have arbitrary signs.  Let $\epsilon$ be an $n$-tuple of
signs, and let $\Lmat^\epsilon$ and $\Mmat^\epsilon$ be the matrices
obtained by replacing $b_j$ with $\epsilon_j b_j$ in the expressions
(\ref{eq:index form of L,M})
 for $\Lmat$ and $\Mmat$.  Thus, for
$\epsilon = (1,\ldots,1)$, $\Lmat^\epsilon = \Lmat$, while for
$\epsilon = (1,\ldots,1,-1)$, $\Lmat^\epsilon = \Lmatbar$. We regard
$\epsilon$ as $n$-periodic in its index.
\begin{prop}\label{prop: Poisson bracket of spectral components of L}
  Let $\uvec$ and $\vvec$ denote eigenvectors
  of $\Lmat^\epsilon_*$ with the same eigenvalue $\lambda$, ie
  \begin{equation}
    \label{eq:uvec_and_vvec}
    \Lmat^\epsilon_*\cdot \uvec = \lambda \uvec, \quad
    \Lmat^\epsilon_*\cdot \vvec = \lambda \vvec,
  \end{equation}
where $\uvec$ and  $\vvec$ need not be linearly independent.   Similarly, let
$\wvec$ and $\xvec$ denote eigenvectors of $\Lmat^\sigma$ with  the same
eigenvalue $\mu$,
 \begin{equation}
    \label{eq:wvec_and_xvec}
    \Lmat^\sigma_*\cdot \wvec = \mu \wvec, \quad
    \Lmat^\sigma_*\cdot \xvec = \mu \xvec.
  \end{equation}
Let 
 \begin{align}
   \label{eq:L_uv,etc}
   L^\epsilon_{uv} &= \uvec \cdot \Lmat^\epsilon \cdot \vvec,\nonumber\\
   L^\sigma_{wx}& = \wvec \cdot \Lmat^\epsilon \cdot \xvec.
 \end{align}
%%  \begin{equation}
%%    \label{eq:L_uv,etc}
%%    L^\epsilon_{uv} = \uvec \cdot \Lmat^\epsilon \cdot \vvec,\quad
%%    L^\sigma_{wx} = \wvec \cdot \Lmat^\epsilon \cdot \xvec.
%%  \end{equation}
Then if $\lambda = \mu$ and $ \prod_{m=1}^n \epsilon_m \sigma_m  = 1$,
  \begin{equation}
    \label{eq:nonzero pbs_of_spectral_comps}
    \{L^\epsilon_{uv}, L^\sigma_{wx}\}_* =      \frac{1}{n} 
\left(
(\vvec\cdot \Dmat\cdot \xvec) (\uvec \cdot \Mmat^\epsilon_* \Dmat\cdot
      \wvec) +
(\uvec\cdot \Dmat\cdot \wvec) (\vvec \cdot \Mmat^\epsilon_* \Dmat\cdot
      \xvec)  
\right),
  \end{equation}
where $\Dmat$ is the diagonal matrix with diagonal elements 
\begin{equation}
  \label{eq:d_m}
  d_m = \epsilon_1 \sigma_1 \cdots \epsilon_{m-1}\sigma_{m-1}.
\end{equation}
Otherwise, ie if $\lambda \ne \mu$ or $ \prod_{m=1}^n \epsilon_m
\sigma_m  = -1$, then
\begin{equation}
  \label{eq:zero_pbs}
   \{L^\epsilon_{uv}, L^\sigma_{wx}\}_* = 0.
\end{equation}
\end{prop}
We note that $\prod_{m=1}^n \epsilon_m \sigma_m  = 1$ if
and only if $\Lmat^\epsilon$ and $\Lmat^\sigma$ are conjugate.  In
fact, we shall only use the cases where $\Lmat^\epsilon$ and
$\Lmat^\sigma$ are (independently) either $\Lmat$ or $\Lmatbar$.

\begin{proof}
%  Given arbitrary vectors $\uvec$, $\vvec$, $\wvec$ and $\xvec$ (not
%  necessarily eigenvectors of $\Lmat^\epsilon$ and $\Lmat^\sigma)$),
%  it is 
%  straightforward to show, 
Using (\ref{eq:index form of L,M}) 
and 
(\ref{eq:pb b_r,p_r}), it is straightforward to show that
%that $\{L^\epsilon_{uv}, ^\sigma_{wx}\}$ is
%  given by
  \begin{align}\label{eq: PB-step-1}
     \{L^\epsilon_{uv}, L^\sigma_{wx}\} =
  &\half \sum_{m=1}^n v_m x_m (A_m(\uvec,\wvec) +
     \epsilon_{m-1}\sigma_{m-1} A_{m-1}(\uvec,\wvec))+\nonumber\\
  &\half \sum_{m=1}^n u_m w_m (A_m(\vvec,\xvec) +
     \epsilon_{m-1}\sigma_{m-1} A_{m-1}(\xvec,\vvec)),
  \end{align}
where
\begin{align}
  \label{eq:A_m}
  A_m(\uvec,\wvec) &= b_m(\epsilon_m u_{m+1}w_m - \sigma_m u_m
  w_{m+1}),\nonumber\\
  A_m(\vvec,\xvec) &= b_m(\epsilon_m v_{m+1}x_m - \sigma_m v_m x_{m+1}).
\end{align}
%and similarly for $A_m(\vvec,\xvec)$.  
Indeed, (\ref{eq: PB-step-1}) holds independently of
the eigenvector equations
(\ref{eq:uvec_and_vvec}) and (\ref{eq:wvec_and_xvec}).
%For convenience, we regard
%$A_m$ as $n$-periodic in its index.
The eigenvector equations imply additionally that, at $\zvec_*$,
$A_{m*}(\uvec,\wvec)$ and $A_{m*}(\vvec,\xvec)$ 
satisfy a Wronskian-like first-order difference
 equation.  Indeed,  (\ref{eq:uvec_and_vvec})
and (\ref{eq:wvec_and_xvec}) yield second-order difference
equations for $\uvec$ and $\wvec$
%in analogy with the difference equation 
(cf (\ref{eq:diff eqn})), 
\begin{subequations}
\begin{align}
\left(   \epsilon_m b_{m} u_{m+1} + p_{m} u_m + 
\epsilon_{m-1}b_{m-1}u_{m-1}\right)_* &= \lambda u_m,\label{eq:
   diff eqn for Lepsilon}\\
\left(   \sigma_m b_{m} w_{m+1} + p_{m} w_m 
+ \sigma_{m-1}b_{m-1}w_{m-1}\right)_* 
&= \mu w_m.\label{eq: diff eqn for Lsigma}
\end{align}
\end{subequations}
Multiplying (\ref{eq: diff eqn for Lepsilon}) by $w_m$ and (\ref{eq:
  diff eqn for Lsigma}) by $u_m$, and subtracting, we get
\begin{subequations} \label{eq:eigenvec difference eqns}
\begin{equation}
  \label{eq:A_m-difference-eqn-uw}
  A_{m*}(\uvec,\wvec) - \epsilon_{m-1}\sigma_{m-1}  A_{m-1*}(\uvec,\wvec)
  = (\lambda-\mu)u_m w_m.
\end{equation}
Similarly,
\begin{equation}
  \label{eq:A_m-difference-eqn-vx}
  A_{m*}(\vvec,\xvec) - \epsilon_{m-1}\sigma_{m-1}  A_{m-1*}(\vvec,\xvec)
  = (\lambda-\mu)v_m x_m.
\end{equation}
\end{subequations}

Suppose that $\lambda \ne \mu$.  From (\ref{eq:eigenvec difference
  eqns}), we have that
\begin{subequations}
\begin{align}
  u_m w_m &= \frac{1}{\Delta}(A_m(\uvec,\wvec) -
  \epsilon_{m-1}\sigma_{m-1}  A_{m-1}(\uvec,\wvec))_*,\\
  v_m x_m &= \frac{1}{\Delta}(A_m(\vvec,\xvec) - \epsilon_{m-1}\sigma_{m-1}  A_{m-1}(\vvec,\xvec))_*,
\end{align}
\end{subequations}
where $\Delta = \lambda - \mu$.  Substituting into (\ref{eq:
  PB-step-1}), we get that
\begin{multline}
    \{L^\epsilon_{uv}, L^\sigma_{wx}\}_* =
    \frac{1}{2\Delta}\sum_{m=1}^n\Big [\\
 \left(A_m(\vvec,\xvec) - \epsilon_{m-1}\sigma_{m-1} A_{m-1}(\vvec,\xvec)\right)\times
\left(A_m(\uvec,\wvec) + \epsilon_{m-1}\sigma_{m-1}
  A_{m-1}(\uvec,\wvec)\right) +\\
\left(A_m(\uvec,\wvec) - \epsilon_{m-1}\sigma_{m-1} A_{m-1}(\uvec,\wvec)\right)\times
\left(A_m(\vvec,\xvec) + \epsilon_{m-1}\sigma_{m-1}
  A_{m-1}(\vvec,\xvec)\right) 
\Big ]_*  \\
=  \frac{1}{\Delta}\sum_{m=1}^n \Big [ 
A_m(\uvec,\wvec) A_m(\vvec,\xvec)  -  A_{m-1}(\uvec,\wvec) A_{m-1}(\vvec,\xvec)
\Big ]_* = 0,
\end{multline}
as required by (\ref{eq:zero_pbs}).

Next, suppose that $\lambda = \mu$.  From 
(\ref{eq:eigenvec difference eqns}),
\begin{subequations}\label{eq: lambda = mu}
\begin{align}
  A_{m*}(\uvec,\wvec) &= \epsilon_{m-1}\sigma_{m-1}
  A_{m-1*}(\uvec,\wvec),\label{eq: l=u,uw}\\
  A_{m*}(\vvec,\xvec) &= \epsilon_{m-1}\sigma_{m-1}
  A_{m-1*}(\vvec,\xvec)\label{eq: l=u, vx}.  
\end{align}
\end{subequations}
Substituting into (\ref{eq: PB-step-1}), we get that
\begin{equation}\label{eq: PB-step-2}
     \{L^\epsilon_{uv}, L^\sigma_{wx}\}_* =
    \sum_{m=1}^n v_m x_m A_{m*}(\uvec,\wvec)+
  \sum_{m=1}^n u_m w_m A_{m*}(\vvec,\xvec).
\end{equation}
Iterating the relations 
(\ref{eq: lambda = mu})
$n$ times, we get that
$A_{m*}(\uvec,\wvec) = \epsilon_{1}\sigma_{1}\cdots
\epsilon_{n}\sigma_{n} A_{m*}(\uvec,\wvec)$ and similarly for
$A_{m*}(\vvec,\xvec)$.  Therefore, if $\epsilon_{1}\sigma_{1}\cdots
\epsilon_{n}\sigma_{n} = -1$,  $A_{m*}(\uvec,\wvec) =
A_{m*}(\vvec,\xvec) = 0$, so that $\{L^\epsilon_{uv},\, L^\sigma_{wx}\}_* =
0$, as in (\ref{eq:zero_pbs}).

Now suppose that  $\epsilon_{1}\sigma_{1}\cdots
\epsilon_{n}\sigma_{n} = 1$.  
In this case, (\ref{eq: l=u,uw})
implies that 
\begin{equation}
  \label{eq:dm Am constant}
  d_m A_{m*}(\uvec,\wvec) = d_{m-1} A_{m-1*}(\uvec,\wvec),
\end{equation}
where $d_m$ is given by (\ref{eq:d_m}), ie, $d_m
A_{m}(\uvec,\wvec) $ is independent of $m$.
Thus, the first term in (\ref{eq: PB-step-2})
can be expressed as 
\begin{multline}
  \label{eq:PB-step-3}
   \sum_{m=1}^n v_m x_m A_{m*}(\uvec,\wvec) = \left(\sum_{m=1}^n d_m v_m
     x_m\right) 
\left(\frac{1}{n} \sum_{m=1}^n
     d_m A_{m*}(\uvec,\wvec)\right) \\ = (\vvec\cdot\Dmat\cdot\xvec) 
\left(\frac{1}{n} \sum_{m=1}^n
     d_m A_{m*}(\uvec,\wvec)\right).
\end{multline}
The sum in the last term in (\ref{eq:PB-step-3})  can be expressed as
\begin{multline}
  \label{eq:PB-step-4}
   \sum_{m=1}^n d_m A_{m*}(\uvec,\wvec) = \sum_{m=1}^n d_m
b_{m*}(\epsilon_m u_{m+1}w_m - \sigma_m u_m w_{m+1}) = \\
\sum_{m = 1}^n \epsilon_m b_{m*} (u_{m+1}(\Dmat\cdot \wvec)_m -  u_m
(\Dmat\cdot \wvec)_{m+1}) = \uvec\cdot\Mmat^\epsilon_* \Dmat\cdot \wvec.
\end{multline}
Substituting (\ref{eq:PB-step-3}) and (\ref{eq:PB-step-4}) into the
first term of
(\ref{eq: PB-step-2}) and making analogous substitutions for the second
term, we get 
\begin{equation}
  \label{eq:PB-final}
   \{L^\epsilon_{uv}, L^\sigma_{wx}\}_* = 
 \frac{1}{n} (\vvec\cdot \Dmat\cdot \xvec) (\uvec \cdot \Mmat^\epsilon_* \Dmat\cdot
      \wvec) + \frac{1}{n} 
(\uvec\cdot \Dmat\cdot \wvec) (\vvec \cdot \Mmat^\epsilon_* \Dmat\cdot
      \xvec),
\end{equation}
as in (\ref{eq:nonzero pbs_of_spectral_comps}).
\end{proof}
\begin{prop}\label{prop: xi,eta canonical}
Let $\xi_r$, $\eta_r$, $\xibar_s$ and $\etabar_s$ be given by
(\ref{eq: xi,eta}).  Then
\begin{subequations}\label{eq: PBs for xi,eta,xibar,etabar}
    \begin{gather}
    \{\xi_r, \xibar_s\}_* =    
    \{\xi_r, \etabar_s\}_* =     
    \{\eta_r, \xibar_s\}_* =    
    \{\eta_r, \etabar_s\}_* =   0, 
\label{eq: PB xi xibar}\\
   \{\xi_r, \xi_t\}_* =    
   \{\eta_r, \eta_t\}_* =     
   \{\xibar_s, \xibar_u\}_* =    
   \{\etabar_s, \etabar_u\}_* = 0,\label{eq: PB xi xi}\\
\{\xi_r,\eta_t\}_* = \frac{1}{n} 
\uvec_{2r}\cdot \Mmat_* \cdot \uvec_{2r-1} \,\delta_{rt},\ \
\{\xibar_s,\etabar_u\}_* = \frac{1}{n} 
\uvecbar_{2s}\cdot \Mmatbar_* \cdot \uvecbar_{2s-1} \,\delta_{su}\label{eq: PB xi eta}.
  \end{gather}
\end{subequations}
Moreover, $\uvec_{2r}\cdot \Mmat_* \cdot \uvec_{2r-1}$ and
$\uvecbar_{2s}\cdot \Mmatbar_* \cdot \uvecbar_{2s-1}$ do not vanish.
\end{prop}
  \begin{proof}
We treat a representative case for each set of relations in
    (\ref{eq: PBs for xi,eta,xibar,etabar}).  For~(\ref{eq: PB xi xibar}),
we consider $\{\eta_r, \xibar_s\}_*$.  From (\ref{eq: xi,eta}) and (\ref{eq:Lambda elements}),
\begin{multline}
  \label{eq:d eta_r}
  d\eta_{r*} = d(\uvec_{2r-1}\cdot \Rmat^T\Lmat\Rmat\cdot \uvec_{2r})_* = 
\uvec_{2r-1}\cdot [\Lmat, d\Rmat]_*\cdot \uvec_{2r} +
 \uvec_{2r-1}\cdot d\Lmat_* \cdot \uvec_{2r}
 \\
= \uvec_{2r-1}\cdot d\Lmat_*\cdot \uvec_{2r},
\end{multline}
where we have used the fact that $\Rmat_* = \Imat$, $d\Rmat_*$ is antisymmetric
(since $\Rmat$ is orthogonal) and, in the last equality, that
$\uvec_{2r-1}$ and $\uvec_r$ are eigenvectors of $\Lmat_*$ with the
same eigenvalue .  Similarly, from (\ref{eq: xi,eta}) and (\ref{eq:Lambdabar elements}),
\begin{equation}
  \label{eq:d xibar_s}
  d\xibar_{s*} = \half \uvecbar_{2s}\cdot d\Lmatbar_*\cdot \uvecbar_{2s}
-  \half \uvecbar_{2s-1}\cdot d\Lmatbar_*\cdot \uvecbar_{2s-1}.
\end{equation}
Together, (\ref{eq:d eta_r}) and (\ref{eq:d xibar_s}) imply that
\begin{equation}
  \label{eq:eta,etabar in terms of L_uv}
   \{\eta_r, \xibar_s\}_* = 
\{ \uvec_{2r-1}\cdot \Lmat \cdot \uvec_{2r},
\half \uvecbar_{2s}\cdot \Lmatbar\cdot \uvecbar_{2s} - 
 \half \uvecbar_{2s-1}\cdot \Lmatbar\cdot \uvecbar_{2s-1} \}_*.
\end{equation}
Proposition~\ref{prop: Poisson bracket of spectral components of L} then implies, with $\epsilon =
(1,\ldots,1)$ and $\sigma = (1,\ldots,-1)$, that
\begin{equation}
  \label{eq:xi_xibar_result}
  \{\eta_r, \xibar_s\}_* = 0,
\end{equation}
since $\prod_{m=1}^n \epsilon_m \sigma_m = -1$.
The remaining relations in (\ref{eq: PB xi xibar}) are obtained
similarly.

For (\ref{eq: PB xi xi}), we consider $\{\eta_r,\eta_t\}_*$.  If $r =
t$ the bracket obviously vanishes, so we take $r \ne t$.  From
(\ref{eq:d eta_r}) 
it follows that
\begin{equation}
  \label{eq:eta,eta in terms of L_}
  \{\eta_r,\eta_t\}_* = 
\{ \uvec_{2r-1}\cdot \Lmat\cdot \uvec_{2r},
   \uvec_{2t-1}\cdot \Lmat\cdot \uvec_{2t}\}_*.
\end{equation}
Since $\lambda_r \ne \lambda_t$, Proposition~\ref{prop: Poisson
  bracket of spectral components of L}
implies that $\{\eta_r,\eta_t\}_* = 0$.  The remaining relations in (\ref{eq:
  PB xi xi}) are obtained similarly.

For (\ref{eq: PB xi eta}), we consider $\{\xi_r,\eta_t\}_*$.
From (\ref{eq:d eta_r}) and (\ref{eq:d xibar_s}) (or rather, its
analog for $d\xi_r$), we get that
\begin{equation}
  \label{eq:xi,eta in terms of L_}
  \{\xi_r,\eta_t\}_* = 
\half \{ \uvec_{2r}\cdot \Lmat \cdot \uvec_{2r}, \uvec_{2t-1}\cdot
\Lmat\cdot \uvec_{2t}\}_*
-
\half \{ \uvec_{2r-1}\cdot \Lmat \cdot \uvec_{2r-1}, \uvec_{2t-1}\cdot
\Lmat\cdot \uvec_{2t}\}_*.
\end{equation}
Proposition~\ref{prop: Poisson bracket of spectral components of L}
then implies, with $\epsilon = \sigma = (1,\ldots,1)$, that
(\ref{eq:xi,eta in terms of L_}) vanishes if $r \ne t$ (as, in this
case, $\lambda_r \ne \lambda_t$).  On the other hand, for $r = t$,
Proposition~\ref{prop: Poisson bracket of spectral components of L},
with $\Dmat = \Imat$, and $\uvec_{2r-1}$, $\uvec_{2r}$ orthonormal
gives that
\begin{equation}
  \label{eq:xi eta result}
   \{\xi_r,\eta_r\}_* = 
\frac{1}{2n} \uvec_{2r}\cdot \Mmat_* \cdot \uvec_{2r-1} - 
\frac{1}{2n} \uvec_{2r-1}\cdot \Mmat_* \cdot \uvec_{2r} = \frac{1}{n} 
\uvec_{2r-1}\cdot \Mmat_* \cdot \uvec_{2r}.
\end{equation}
The quantity $\uvec_{2r}\cdot \Mmat_* \cdot \uvec_{2r-1}$ which appears
in (\ref{eq:xi eta result}) does not vanish.  This is because,
from (\ref{eq:dm Am constant}) and (\ref{eq:PB-step-4}) (with
$d_m = 1$), 
we have that $\uvec_{2r}\cdot \Mmat_* \cdot \uvec_{2r-1}$ is equal to 
$n b_m (u_{2r-1,m+1} u_{2r,m} - u_{2r-1,m} u_{2r,m+1})$ for all $m$.
$\uvec_{2r}\cdot \Mmat_* \cdot \uvec_{2r-1} = 0$ would imply
that $\uvec_{2r-1}$ and $\uvec_{2r}$ are
proportional, contradicting orthonormality. The result for
$\{\xibar_s,\etabar_u\}_*$ follows similarly.
  \end{proof}

\begin{thm}\label{thm: symplectic submanifold}
$\Sigma_k$ is a codimension-$2k$ symplectic submanifold.  It consists
of ${n-1 \choose k}$ components which are disconnected from each
other.  $\Sigma_k$ is contained in the closure of $\Sigma_j$ for all
$j < k$.
\end{thm}
\begin{proof}
  Let $\zvec_* \in \Sigma_k$ and let $\nu_*$, $\nubar_*$ denote the
  number of degenerate eigenvalues of $\Lmat_*$ and $\Lmatbar_*$,
  respectively, so that, by Theorem~\ref{thm: sing and degen}, $\nu_*
  + \nubar_* = k$.  Let $\xi_r$, $\eta_r$, $1 \le r \le \nu_*$, and
  $\xibar_s$, $\etabar_s$, $1 \le s \le \nubar_*$, be given by
  (\ref{eq: xi,eta}) in some neighbourhood of $\zvec_*$.
%%  (\ref{eq:local_Sigma_k}), I
In this
  neighbourhood, 
$\Sigma_k$ is given by 
   \begin{equation}
     \label{eq:zero_set}
     \xi_r = \eta_r = \xibar_s = \etabar_s = 0.
   \end{equation}
Proposition~\ref{prop: xi,eta canonical} implies that the derivatives of $\xi_r$, $\eta_r$,
$\xibar_s$ and $\etabar_s$ are linearly independent at $\zvec_*$.  It
follows from the 
implicit function theorem that $\Sigma_k$ is codimension-$2k$ submanifold.

Next, we show that the tangent space of $\Sigma_k \subset \Rr^{2n}$ at
$\zvec_*$ is symplectic.
%is skew-orthogonal to the subspace $E_* \subset \Rr^{2n}$
%spanned by the Hamiltonian vector fields generated by $\xi_r$,
%$\eta_r$, $\xibar_s$, $\etabar_s$ at $\zvec_*$.  
Let $\Xvec_*\in
T_{\zvec_*}\Sigma_k$.  $\Xvec_*$ may be extended to a vector field
$\Xvec$ defined in a neighbourhood of $\zvec_*$ whose restriction to
$\Sigma_k$ is tangent to $\Sigma_k$.  Then (\ref{eq:zero_set})
implies, for example, that $(\Xvec\contract d\xi_r)_* = 0$.  But
$(\Xvec\contract d\xi_r)_*$ is just the symplectic inner product of
$\Xvec_*$ with the Hamiltonian vector field generated by $\xi_r$ at
$\zvec_*$.  Let $E_*$ denote the subspace of $T_*\Rr^{2n}$ spanned by
the Hamiltonian vector fields generated by $\xi_r$, $\eta_r$, and
$\xibar_s$, $\etabar_s$.  Arguing as above, we may conclude that
$\Xvec_*$, and therefore $T_*\Sigma_k$, 
is skew-orthogonal to $E_*$.
Proposition~\ref{prop: xi,eta canonical}
implies that $E_*$ is symplectic of dimension $2k$.  Since
$\codim T_{\zvec_*}\Sigma_k = 2k$, it follows that
$T_{\zvec_*}\Sigma_k$ is the skew-orthogonal complement of $E_*$.  The
skew-orthogonal complement of a symplectic subspace is itself
symplectic, so $T_{\zvec_*}\Sigma_k$ is symplectic, and therefore
$\Sigma_k$ is a symplectic submanifold.

$\Sigma_k$ can be partitioned into distinct components according to the particular $k$ pairs of
eigenvalues which are degenerate.  From Proposition~\ref{prop: allowed degens}
there are $(n-1)$ possible degeneracies
($[\half(n-1)]$ from $\Lmat$ and $[\half n]$ from $\Lmatbar$),
so the maximum number of
components is ${n-1 \choose k}$.  Every choice of $k$ pairs can
be realised by setting, in a neighbourhood of a point of
$\Sigma_{n-1}$, the appropriate pairs of coordinates $(\xi_r,
\eta_r)$, $(\xibar_s,\etabar_s)$ to be nonzero, and the remaining 
pairs to be nonzero.  Thus $\Sigma_k$ has precisely ${n-1 \choose k}$
components.  Along any path connecting points in different components,
the number of eigenvalue degeneracies must change, so the path 
must leave
$\Sigma_k$.  Therefore, the components are disconnected from
each other.

Any neighbourhood of $\zvec_* \in \Sigma_k$ contains points of
$\Sigma_{j<k}$.  Such points are obtained by setting $k - j$ of the
$k$ coordinates pairs $(\xi_r, \eta_r)$ and $(\xibar_s,
\etabar_s)$ to be nonzero.  Therefore, $\Sigma_k$ is contained in the
closure of $\Sigma_{j < k}$.
\end{proof}

%% Any neighbourhood of $\zvec_* \in \Sigma_k$ contains points of
%% $\Sigma_{j<k}$.  Such points are obtained by setting $k - j$
%% of the pairs $(\xi_r, \eta_r)$ and $(\xibar_s, \etabar_s)$ to be nonzero.
%% Since $\Sigma_{n-1}$ is nonempty (cf Proposition~\ref{prop: Omega_n}),
%% $\Sigma_k$ is nonempty for $k < n$, and $\Sigma_k$ is
%% contained in the closure of $\Sigma_{j < k}$.

We consider next the transverse stability of $\Sigma_k$.  Let $\zvec_*
\in \Sigma_k$.  In analogy with (\ref{eq: xi,eta}), we define the
local functions
\begin{align}\label{eq: tau_r}
  \tau_r &= \half (\Lambda_{2r,2r} + \Lambda_{2r-1,2r-1}),
&1 \le r \le \nu,\nonumber\\
\taubar_s &= \half (\Lambdabar_{2s,2s} + \Lambdabar_{2s-1,2s-1}),
&1 \le s \le \nubar,
\end{align}
with
\begin{align}\label{eq: tau_r*}
  \tau_{r*} &= \half \uvec_{2r}\cdot d\Lmat_*\cdot \uvec_{2r}
+  \half \uvec_{2r-1}\cdot d\Lmat_*\cdot \uvec_{2r-1},&1 \le r \le \nu,\nonumber\\
\taubar_{s*} &=  \half \uvecbar_{2s}\cdot d\Lmatbar_*\cdot \uvecbar_{2s}
+  \half \uvecbar_{2s-1}\cdot d\Lmatbar_*\cdot \uvecbar_{2s-1}, &1 \le s \le \nubar.
\end{align}
With arguments similar to those of Proposition
\ref{prop: xi,eta
  canonical},
one can show that  
the Poisson brackets of the $\tau_r$'s and the $\taubar_s$'s
with $\xi_r$, $\eta_r$, $\xibar_s$ and $\etabar_s$ all vanish at $\zvec_*$.  Likewise, the
Poisson brackets of the $\tau_r$'s and $\taubar_s$'s amongst themselves
vanish at $\zvec_*$.  We record this briefly as
\begin{equation}
  \label{eq:C brackets}
  \{\tau_{\cdot}, \chi_{\cdot}\}_* = \{\tau_{\cdot}, \tau_{\cdot\cdot}\}_* = 0,
\end{equation}
where $\chi_{\cdot}$ denotes  $\xi_r$, $\eta_r$, $\xibar_s$ or
$\etabar_s$, and $\tau_{\cdot}$, $\tau_{\cdot\cdot}$ denotes $\tau_r$ or $\taubar_s$.

Let
\begin{align}
  \label{eq:P_r,Pbar_s}
  T_r(x) &= \frac{\det(\Lmat_* - x \Imat)}{ \lambda_{r*} - x} 
  = \sum_{j=1}^{n}c_{rj}x^{j-1}, 
  &1\le r \le \nu,\nonumber\\ 
  \Tbar_{s*}(x) &= \frac{\det(\Lmatbar_* - x \Imat)}
{\lambdabar_{s*}-x }
  = \sum_{j=1}^{n}\cbar_{sj}x^{j-1}, 
  &1\le s \le \nubar.
\end{align}
Clearly $T_r(\Lmat_*) = 0$, and moreover, the $T_r$'s are linearly
independent (since, if $\sum_{t=1}^\nu a_{t} T_{t}(\lambda_{r*}) = 0$,
then $\sum_{t=1}^\nu a_{t} T'_{t}(\lambda_{r*}) = 0$, and the fact that
$T'_t(\lambda_{r_*}) = 0$ for $r\ne t$ implies that $a_r = 0$ for
each $r$).
Therefore, the $T_r$'s constitute a basis for $\Tcal_*^n$, the space
of polynomials of degree at most $n$ that annihilate $\Lmat_*$ (cf
(\ref{eq:Tcal^n})).  Likewise, the $\Tbar_s$'s constitute a basis for
$\Tbarcal_*^n$, the space of polynomials of degree at most $n$ that
annihilate $\Lmatbar_*$ (cf (\ref{eq:Tbarcal_*^n})).  It follows from
Corollary~\ref{cor: t_* and V_*} that the vectors $\cvec_r =
(c_{r1},\ldots,c_{rn})$ and $\cvecbar_s =
(\cbar_{s1},\ldots,\cbar_{sn})$ constitute a basis for $V_*$, the
space of linear relations amongst the $dF_{j*}$'s. Therefore, letting
\begin{equation}\label{eq: G,Gbar}
  G_r = \sum_{j=1}^n c_{rj} F_j,\quad
  \Gbar_s = \sum_{j=1}^n \cbar_{sj} F_j,
\end{equation}
we have that the Hamiltonian flows generated by $G_r$ and $\Gbar_s$ have $\zvec_*$ as
a fixed point.  The stability of these flows at $\zvec_*$ is given by
the following:
\begin{thm}\label{thm: transverse stability}
\begin{align}\label{eq:G_r''-statement of result}
   G''_{r*}     &= 2T'_r(\lambda_{r*}) \left(
d\xi_r\otimes d\xi_r +
d\eta_r\otimes d\eta_r + 
d\tau_r \otimes d\tau_r
\right)_*,\nonumber\\
   \Gbar''_{s*}     &= 2\Tbar'_s(\lambdabar_{s*}) \left(
d\xibar_s\otimes d\xibar_s +
d\etabar_s\otimes d\etabar_s + 
d\taubar_s \otimes d\taubar_s
\right)_*.
\end{align}
Thus, the linearised $G_r$-flow about $\zvec_*$ produces elliptic
oscillations 
in the $(\xi_r,\eta_r)$ plane with frequencies 
\begin{equation}
  \label{eq:omega_r}
  \omega_r = 2n T'(\lambda_{r*})/(\uvec_{2r}\cdot\Mmat_*\cdot\uvec_{2r-1}).
\end{equation}
Similarly,
the linearised $\Gbar_s$-flow about $\zvec_*$ produces elliptic oscillations
in the $(\xibar_s,\etabar_s)$ plane with frequencies
\begin{equation}
  \label{eq:omegabar_s}
  \omegabar_s = 2n \Tbar'(\lambdabar_{s*})/(\uvecbar_{2s}\cdot\Mmatbar_*\cdot\uvecbar_{2s-1}).
\end{equation}
\end{thm}
\begin{proof}
  We carry out the calculations for $G_r$; those for $\Gbar_s$ are similar.
From (\ref{eq: G,Gbar}) and (\ref{eq:F_j}), we have that
  \begin{equation}
    \label{eq:G''_r-first}
    G''_{r*} = \sum_{j=1}^n \Tr (c_{jr} \Lmat_*^{j-1}\Lmat''_*) +
    \sum_{j=1}^n c_{jr} \Tr(d\Lmat^{j-1}\otimes d\Lmat)_*.
  \end{equation}
As $T_r(\Lmat_*) = 0$, the first term vanishes.  As for the second term,
\begin{align}\label{eq: second term in G''_r}
   \sum_{j=1}^n c_{jr} \Tr(d\Lmat^{j-1}\otimes d\Lmat)_* &=
     \sum_{j=1}^n \sum_{k=0}^{j-2}  c_{jr} \Tr(\Lmat_*^k d\Lmat_*
     \Lmat_*^{j-k-2}\otimes d\Lmat_*) \nonumber\\
&= \sum_{a=1}^{n-\nu} \sum_{b=1}^{n-\nu}  \sum_{j=1}^n
     \sum_{k=0}^{j-2}  c_{jr} \lambda_{a*}^k \lambda_{b*}^{j-k-2} \Tr(\rho_{a*}
     d\Lmat_*\otimes \rho_{b*} d\Lmat_*),
\end{align}
where we have introduced the spectral resolution of $\Lmat_*$,
\begin{equation}
  \label{eq:spectral resolution}
  \Lmat_* = \sum_{a=1}^{n-\nu} \lambda_{a*} \rho_{a*}, \quad
%  \label{eq:rho_a}
  \rho_{a*} = 
\frac{\prod_{r\ne a} (\Lmat_* - \lambda_{a*})}{ \prod_{r\ne a}(\lambda_{r*} - \lambda_{a*})}.
\end{equation}
$\rho_{a*}$ is the symmetric projector onto the $\lambda_{a*}$-eigenspace of
$\Lmat_{a*}$, so that $\rho_{a*} \rho_{b*} = \delta_{ab} \rho_{a*}$.  The sums over
$j$ and $k$ in (\ref{eq: second term in G''_r}) are readily performed to give
\begin{equation}
  \label{eq:j,k sum}
   \sum_{j=1}^n
     \sum_{k=0}^{j-2}  c_{jr} \lambda_{a*}^k \lambda_{b*}^{j-k-2} = 
     \begin{cases}
       (T_r(\lambda_{a*}) - T_r(\lambda_{b*}))/(\lambda_{a*} - \lambda_{b*}),&
       a \ne b,\\
       T_r'(\lambda_{a*}),& a = b.
     \end{cases}
\end{equation}
We have that $T_r(\lambda_{a*}) = 0$ for all $a$, while $T_r'(\lambda_{a*})$
vanishes unless $a = r$.  Thus, the sums over $a$ and $b$ in (\ref{eq:
  second term in G''_r}) collapse to the single term $a = b = r$.
Substituting into (\ref{eq:G''_r-first}), we get 
\begin{equation}
  \label{eq:G_r''-second}
  G_{r*}'' = T_r'(\lambda_{r*}) \Tr(\rho_{r*} d\Lmat_*\otimes \rho_{r*} d\Lmat_*).
\end{equation}
The projector $\rho_{r*}$ may be written as 
the diadic $\uvec_{2r-1} \uvec_{2r-1} +  \uvec_{2r}\uvec_{2r}$.
%% \begin{equation}
%%   \label{eq:rho_r}
%%   \rho_{r*} = \uvec_{2r-1} \uvec_{2r-1} +  \uvec_{2r}\uvec_{2r}.
%% \end{equation}
Substituting into (\ref{eq:G_r''-second}), we get
\begin{multline}
  \label{eq:G_r''-third}
  G_{r*}'' = T_r'(\lambda_{r*}) \left[(\uvec_{2r}\cdot d\Lmat_*\cdot \uvec_{2r}) \otimes
  (\uvec_{2r}\cdot d\Lmat_*\cdot \uvec_{2r}) \right.\\
 + 2(\uvec_{2r-1}\cdot d\Lmat_*\cdot
  \uvec_{2r})\otimes (\uvec_{2r-1}\cdot d\Lmat_*\cdot \uvec_{2r})\\ 
\left. + 
(\uvec_{2r-1}\cdot d\Lmat_*\cdot \uvec_{2r-1}) \otimes (\uvec_{2r-1}\cdot d\Lmat_*\cdot \uvec_{2r-1})\right].
\end{multline}
Since (cf (\ref{eq:d eta_r}), (\ref{eq:d xibar_s}), (\ref{eq: tau_r*}))
\begin{align}
  \label{eq:dxis,etc}
  d\xi_{r*} +  d\tau_{r*} &=  \uvec_{2r}\cdot d\Lmat_*\cdot \uvec_{2r},\nonumber\\
 d\xi_{r*} -  d\tau_{r*} &= \uvec_{2r-1}\cdot d\Lmat_*\cdot
 \uvec_{2r-1},\nonumber\\
d\eta_{r*} &= \uvec_{2r-1}\cdot d\Lmat_*\cdot \uvec_{2r},
\end{align}
(\ref{eq:G_r''-third}) 
yields the
required result (\ref{eq:G_r''-statement of result}). 

 (\ref{eq:G_r''-statement of result}), along with the Poisson
bracket relations of Proposition~\ref{prop: xi,eta canonical} and
those recorded in (\ref{eq:C brackets}), imply that the equations of
motion for the linearised $G_r$-flow about $\zvec_*$ transverse to
$\Sigma_k$ are given by
  \begin{equation}
    \label{eq:linearised G_r}
    \dot \xi_r = \omega_r \eta_r, \ \ \dot \eta_r = -\omega_r \xi_r,
  \end{equation}
while the remaining transverse coordinates $\xi_{u\ne r}, \eta_{u\ne
  r}, \xibar_s, \etabar_s$ are fixed.  Thus, the linearised transverse
  $G_r$-flow is elliptic in the $(\xi_r,\eta_r)$-plane with frequency
  $\omega_r$.  
\end{proof}

%% Since $\Sigma_{n-1}$ is not empty, neither is $\Sigma_k$ for $k <
%% n-1$.  In neighbourhood of $\Sigma_{n-1}$, can break any of the double
%% degeneracies, leaving degeneracies between remaining unbroken ones.
%% More about the global structure?  What is the question one would like
%% to ask?  Are there any more branches?

\section{Maslov indices and eigenvector monodromies.}\label{sec: Maslov}

Consider an integrable system in $\Rr^{2n}$ with integrals
of motion $\Fvec:\Rr^{2n}\rightarrow \Rr^n$.  In the set of regular
points 
$R$ of $\Fvec$, the Hamiltonian vector fields generated by the $F_j$'s
span a $n$-dimensional Lagrangian plane, which we denote
$\lambda(\zvec)$.  As shown by Arnold \cite{arnold}, the space of $n$-dimensional Lagrangian
planes, $\Lambda(n)$, has fundamental group $\pi_1(\Lambda(n)) = \Zz$, and
the Maslov index of a continuous, oriented closed curve $C$ in $R$
is the degree (winding number) of $\lambda(C)$ in $\Lambda(n)$,
\begin{equation}
  \label{eq:mu(C)}
  \mu(C) = \wn \lambda(C).
\end{equation}
For angle contours $C_j$ on an invariant torus, the Maslov index appears in
the semiclassical EBK quantisation conditions  for the associated action
variables \cite{maslov, keller},
\begin{equation}
  \label{eq:EBK}
  I_j = (n_j + \fourth \mu_j) \hbar.
\end{equation}
In this case $\mu(C)$ is always even (this is because the distribution of
Lagrangian planes $\lambda(\zvec)$ over $R$ is orientable).

In a companion paper \cite{FR1}, we show that under certain genericity
conditions, the Maslov index of $C$ can be expressed as a sum of
contributions from the nondegenerate corank-one singularities it encloses.
We briefly summarise the results.  Let $\Sigma$ denote the critical set
of $\Fvec$, and let
\begin{equation}
  \label{eq:Sigma_l}
  \Sigma_k = \{\zvec \in \Rr^{2n} |\, \corank d\Fvec(\zvec) = k\}
\end{equation}
denote the set of critical points of $\Fvec$ of corank $k$.  Given $\zvec_*\in
\Sigma_1$, let $\cvec \in \Rr^n$ be a nontrivial solution of
$\sum_{j=1}^n c_j d\Fvec_{j*}
= 0$ (here $d\Fvec_* = d\Fvec(\zvec_*)$).  Let $K_* = \Jmat\left (\sum_{j=1}^n
c_j  
  \Fmat''_{j*}\right)$. We say that $\zvec_*$ is {\it nondegenerate} if $\Tr
\Kmat_*^2 \ne 0$.  Let $\Delta$ denote the set of nondegenerate
points in $\Sigma_1$,
\begin{equation}
  \label{eq:Delta}
  \Delta = \{\zvec_* \in \Sigma_1 | \, \Tr \Kmat_*^2 \ne 0 \}.
\end{equation}
In general, $\Delta$ is a codimension-2 symplectic submanifold.
Let $S: D^2 \rightarrow \Rr^{2n}$ denote a
map of the oriented unit two-disk $D^2$ into $\Rr^{2n}$ smooth on the
interior of $D^2$ such
that $S\vert_{\partial D^2} = C$.  We assume that $S$ is {\it
  transverse} to $\Sigma$.  By this we mean that i) the only critical
points of $d\Fvec$ contained in the image of $S$ belong to $\Delta$, ii)
$S^{-1}(\Delta)$ consists of a finite set of points $e_1,\ldots,e_r$
in the interior of $ D^2$, and iii) $dS(e_j)$ has full rank.  (If $\Sigma-\Sigma_1$ is
composed of submanifolds of codimension three or more, such an $S$ can
always be found.)  Let $\zvec_{j} = S(e_j)$ denote the critical points
of $\Fvec$ in the image of $S$.  Let $E_{\zvec_{j}}$ denote the skew-orthogonal
complement of $T_{\zvec_j}\Delta$.  $E_{\zvec_j}$ is a two-dimensional
symplectic plane.  Let $P_j: T_{\zvec_j}\Rr^{2n} \rightarrow E_{\zvec_j}$
denote the projection onto $E_{\zvec_j}$ with respect to the
decomposition
%$ T_{\zvec_j}\Rr^{2n} = T_{\zvec_j}\Delta \oplus E_{\zvec_j}$.
 \begin{equation}
   \label{eq:decomp}
   T_{\zvec_j}\Rr^{2n} = T_{\zvec_j}\Delta \oplus E_{\zvec_j}.
 \end{equation}
Then the map $P_j \circ dS : T_{e_j}D^2 \rightarrow E_{\zvec_j}$
is nonsingular.  Let $\sigma_j = \pm 1$ according to whether this map
is orientation-preserving or reversing (the orientation on
$E_{\zvec_j}$ is given by the symplectic form).  Then
%\begin{thm}\label{thm: thm 3}
  \begin{equation}
    \label{eq:mu(C)2}
    \mu(C) = 2 \sum_j \sigma_j \cdot \sgn \Tr \Kmat_{\zvec_j}^2.
  \end{equation}
%\end{thm}
  
  For the periodic Toda chain, all critical points are nondegenerate
  of elliptic type (cf Theorem~\ref{thm: transverse stability}), so
  that $\Tr \Kmat_{\zvec_j}^2 < 0$.  Also, $\Sigma - \Sigma_1 =
  \cup_{k > 1} \Sigma_k$ is composed of submanifolds of codimension
  four or more (Theorem~\ref{thm: symplectic submanifold}).  Thus, for
  $C$ a continuous closed curve in the regular component of the Toda
  integrals of motion, we have that
\begin{equation}
  \label{eq:todamaslov}
    \mu(C) = -2 \sum_j \sigma_j.
\end{equation}

We can also associate to $C$
the signs acquired by the
normalised eigenvectors of $\Lmat$ and $\Lmatbar$ on continuation
around $C$.  Explicitly, let the eigenvalues of $\Lmat$ and
$\Lmatbar$ be indexed in increasing order, so that $\lambda_1 < \cdots
< \lambda_n$ and $\lambdabar_1 < \cdots < \lambdabar_n$ in $R$.  
Let $\zvec(t)$, $0 \le t \le 1$ denote a parameterisation of $C$, and let
$\uvec_r(t)$ denote a continuously varying, normalised real
eigenvector of $\Lmat(\zvec(t))$ 
with eigenvalue $\lambda_r(\zvec(t))$.  
Then
\begin{equation}
  \label{eq:gamma_r}
  \uvec_r(1) = \gamma_r(C) \uvec_r(0),
\end{equation}
where $\gamma_r(C) = \pm 1$.  Defining $\uvecbar_s(t)$ analogously to be
a continuously varying, normalised eigenvector of $\Lmatbar(\zvec(t))$ with
eigenvalue $\lambdabar_s(\zvec(t))$, we have that
\begin{equation}
  \label{eq:gammabar_s}
  \uvecbar_s(1) = \gammabar_s(C) \uvecbar_s(0),
\end{equation}
where $\gammabar_s(C) = \pm 1$.
$\gamma_r$ and $\gammabar_s$ may be regarded as the
holonomies of the real eigenvector line bundles $E_r$ and $\Ebar_s$
over $R$,
\begin{align}
  \label{eq:E_r, Ebar_s}
  E_r &= \{(\zvec, \uvec) \in R\times \Rr^n |\,
  (\Lmat(\zvec) - \lambda_r(\zvec))\cdot \uvec =
  0\},\nonumber\\
  \Ebar_s &= \{(\zvec, \uvec) \in R\times \Rr^n |\,
  (\Lmatbar(\zvec) - \lambdabar_s(\zvec))\cdot \uvec =
  0\}.
\end{align}
They are examples of (real) geometric phases \cite{berry84}, though in
this context (holonomies of eigenvectors of real symmetric matrices)
have a long history (see, eg, \cite{arnoldcm} and \cite{berry90}).

For the periodic Toda chain, the Maslov index and eigenvector
holonomies are related by the following:
\begin{thm}\label{thm: Maslov and holonomy}
  \begin{equation}
    \label{eq:formula}
    (-1)^{\mu/2} = \prod_{r\ \text{even}}  \gamma_r  \prod_{s\
      \text{even}} \gammabar_s.
  \end{equation}
\end{thm}
We note that the unrestricted product $\prod_r \gamma_r$ is always
$+1$, since this gives the holonomy of the (trivial) determinant
bundle of $R \times \Rr^n$.  Similarly for the unrestricted product
$\prod_s \gammabar_s$.  Thus, either of the products in
(\ref{eq:formula}) may be restricted to odd rather than even indices.
(Alternatively, from Proposition ~\ref{prop: allowed degens} one can deduce that
$\gamma_r = \gamma_{r+1}$ for $r$ even and $\gammabar_{r} =
  \gammabar_{r+1}$ for $r$ odd.)
\begin{proof}
Let $C$ be a continuous closed curve in the regular set $R$ of
$\Fvec$, and let $S$ be a transverse disk with boundary $C$.  Let
$N$ denote the number of singular points in the image of $S$.
From (\ref{eq:todamaslov}), 
\begin{equation}
  \label{eq:(-1)^mu/2}
  (-1)^{\mu(C)/2} = (-1)^N.
\end{equation}
At the singular points $\zvec_j \in \Sigma_1$, there is precisely one doubly degenerate
eigenvalue of either $\Lmat(\zvec_j)$ or $\Lmatbar(\zvec_j)$.  For
definiteness, suppose
$\lambda_r$ and $\lambda_{r+1}$ are degenerate at $\zvec_*$.  As in
(\ref{eq:Lambda elements}) and (\ref{eq:Lambda(zpvec)}), in a
neighbourhood of $\zvec_*$,
$\Lmatbar(\zvec)$ is
smoothly conjugate to a diagonal matrix,
while $\Lmat(\zvec)$ is smoothly conjugate to a block diagonal matrix
$\Lambda(\zvec)$ with a single two-dimensional block,
%, which we denote $\Lambda^{(2)}$,
with elements $\Lambda_{ij}(\zvec)$,
$i,j = 1, 2$, and the rest diagonal.  Let
\begin{equation}
  \label{eq:a,b}
  \xi(\zvec) = \half(\Lambda_{11}(\zvec) - \Lambda_{22}(\zvec)), \quad \eta(\zvec) = \Lambda_{12}(\zvec).
\end{equation}
As in Proposition~\ref{prop:
  xi,eta canonical}, $\xi$ and $\eta$ are functionally independent
  near $\zvec_*$,
and $\xi(\zvec_*) = \eta(\zvec_*) = 0$.
Construct local coordinates $(\xi,\eta,a_1,\ldots,a_{2n-2})$
with origin at $\zvec_*$.  Let $C_*$ denote the closed curve
  parameterised by $\xi(t)
= \epsilon \cos 2\pi t$, $\eta(t) = \epsilon \sin 2\pi t$, $a_k = 0$,
with $\epsilon$ chosen small enough so that $C_*$ lies in the
  coordinate neighbourhood.  Along
$C_*$, let $\Lambda^{(2)}(t)$ denote the two-dimensional block of
  $\Lambda$.  Then
\begin{equation}
  \label{eq:Lambda block} \Lambda^{(2)}(t) =   
\half \tau(t) \Imat + 
  \begin{pmatrix}
  -\cos 2\pi t&\sin 2\pi t\\
\sin 2\pi t&\cos 2\pi t
  \end{pmatrix},
\end{equation}
where $\tau(t)$ is given by $\Lambda_{22} + \Lambda_{11}$ along $C_*$.
The eigenvectors $(\cos \pi t\ \sin\pi t)^T$ and $(\sin \pi t \ 
-\cos \pi t)^T$ of $\Lambda^{(2)}(t)$ change sign around $C_*$.  This
corresponds to holonomies 
\begin{equation}
  \label{eq:-1 holonomies}
  \gamma_r(C_*) = \gamma_{r+1}(C_*) = -1
\end{equation}
in the associated eigenvectors $\uvec_{r}$ and $\uvec_{r+1}$ of $\Lmat$.  The other holonomies, ie
$\gamma_{r\ne p}(C_*)$, $\gammabar_s(C_*)$, are trivially $+1$.

We introduce contours $C_j$ analogously for all singularities
$\zvec_j$ in the image of $S$ (replacing $\Lmat$ by $\Lmatbar$ in the preceding as
appropriate).  $C$ is homologous to an oriented sum of circuits $C_j$.
Around each $C_j$, exactly one even-indexed eigenvector holonomy of
either $\Lmat$ or $\Lmatbar$ is $-1$ (cf (\ref{eq:-1 holonomies})),
and the rest are +1.  Thus,
\begin{equation}
  \label{eq:prod of even hols}
  \prod_{r, even} \gamma_r(C) \prod_{s,even}\gamma_s(C) = (-1)^N.
\end{equation}
Comparing (\ref{eq:prod of even hols}) and (\ref{eq:(-1)^mu/2}), we
obtain the
result (\ref{eq:formula}).
\end{proof}

\section{Discussion}
$\Sigma_k$, the space of codimension-$2k$ singularities of the Toda
chain, is a symplectic submanifold corresponding to points where there
are $k$ doubly degenerate eigenvalues of the Lax matrices $\Lmat$ and
$\Lmatbar$, for $1 \le k \le n-1$.  These submanifolds are of elliptic
type, and we have calculated the frequencies of transverse
oscillations under integrable flows that preserve them pointwise.  The
codimension-two singularities are sources for the Maslov index.  The
(even) Maslov index of a closed curve $C$ is determined, modulo 4, by
the product of the even- (or odd-) indexed eigenvector holonomies of
$\Lmat$ and $\Lmatbar$.  It would be interesting to relate higher
Maslov classes \cite{suzuki, trofimov} to singularities of higher
codimension, and to compute these higher Maslov classes explicitly for
the Toda chain. In this context, it would also be interesting to study
higher-order corrections \cite{littlejohn04, cdv2004} to the
semiclassical quantization conditions~(\ref{eq:EBK}) for the quantum
Toda chain \cite{gutzwiller1, gutzwiller2}.

\bigskip

\noindent{\bf Acknowledgments}\\
We thank the referees for helpful remarks.
JAF was supported by a grant from the EPSRC.  JMR thanks the MSRI for
hospitality and support while some of this work was carried out.
%\section{bibliography}

\bibliography{toda}

\end{document}